\documentclass[12pt,notitlepage,a4paper]{article}

\pdfoutput=1

\usepackage{a4wide}
\usepackage{epsfig}
\usepackage{color,graphicx}
\usepackage{cite}
\usepackage{amssymb,amsmath}

\newcommand{\be}{\begin{equation}}
\newcommand{\ee}{\end{equation}}
\newcommand{\bea}{\begin{eqnarray}}
\newcommand{\eea}{\end{eqnarray}}

\begin{document}

\begin{center}  

\vskip 2cm 

\centerline{\Large {\bf Brane webs and $O5$-planes}}

\vskip 1cm

\renewcommand{\thefootnote}{\fnsymbol{footnote}}

   \centerline{
    {\large \bf Gabi Zafrir${}^{a}$} \footnote{gabizaf@techunix.technion.ac.il}}

\vspace{1cm}
\centerline{{\it ${}^a$ Department of Physics, Technion, Israel Institute of Technology}} \centerline{{\it Haifa, 32000, Israel}}
\vspace{1cm}

\end{center}

\vskip 0.3 cm

\setcounter{footnote}{0}
\renewcommand{\thefootnote}{\arabic{footnote}}   
   
\begin{abstract}

We explore the properties of five-dimensional supersymmetric gauge theories living on $5$-brane webs in orientifold $5$-plane backgrounds. This allows constructing quiver gauge theories with alternating $USp(2N)$ and $SO(N)$ gauge groups with fundamental matter, and thus leads to the existence of new $5d$ fixed point theories. The web description can be further used to study non-perturbative phenomena such as enhancement of symmetry and duality. We further suggest that one can use these systems to engineer $5d$ $SO$ group with spinor matter. We present evidence for this claim.  

 \end{abstract}
 
 \newpage
 
\tableofcontents

\section{Introduction}



Gauge theories in $5d$ are non-renormalizable and thus are not expected to exist as microscopic theories. For example, maximally supersymmetric Yang-Mills theory is believed to flow to the $6d$ $(2,0)$ theory in the UV\cite{Dou,LPS}, so the microscopic theory is actually $6d$. Nevertheless, there is a lot of evidence that in the $\mathcal{N}=1$ supersymmetic case, corresponding to $8$ supercharges, an interacting UV fixed point may exist making the theory UV complete\cite{SEI,SM,SMI}. The gauge theory can then be realized as the IR limit of such a SCFT under a mass deformation, corresponding to the inverse gauge coupling square which has dimension of mass in $5d$. 

An interesting question then is how can we study these $5d$ SCFT's. One way is to embed them in string theory. A convenient embedding is given by $5$-brane webs in type II B string theory\cite{HA,AHK}. This realizes the $5d$ SCFT as an intersection of $5$-branes at a point. The moduli and mass parameters of the SCFT are then realized as motions of the internal and external $5$-branes respectively. In particular, the 5d SCFT may posses a deformation leading to a low-energy gauge theory. 

Thus, $5$-brane webs can be used to study various properties of these theories. First of all they give support for the existence of fixed points for various gauge theories. Not every $5d$ gauge theory flows to a $5d$ SCFT, and the demand that such a SCFT exists is expected to constrain the matter content of the theory. If a gauge theory can be realized as the IR theory in a brane web then this strongly suggests that it flows to a UV fixed point described by the collapsed web. Therefore, brane webs can be used to study the conditions for the existence of fixed points.

Another useful application of brane webs is to study $5d$ dualities. A single SCFT may have more than one gauge theory deformation, in which case these different IR gauge theories are said to be dual. This is somewhat similar to Seiberg duality in $4d$, except that in this case there are several different IR gauge theories all going to the same UV SCFT. This is nicely realized in brane webs, where a SCFT can be deformed in different ways leading to different IR gauge theories \cite{HA}. This usually involves an $SL(2,Z)$ transformation in the brane web. Thus, brane webs provide a useful way to motivate these kind of dualities. For several examples of this see \cite{BPTY,BGZ,Zaf,BZ,BZ1}.

Brane webs can also be used to study symmetry enhancement in $5d$ gauge theories\cite{DHIZ,DHIZ1}. They can be used to calculate the $5d$ superconformal index of a SCFT using the methods of topological strings\cite{IV}. Even when calculating the $5d$ superconformal index for a gauge theory using localization\cite{KKL}, brane webs are very useful for evaluating the instanton contribution\cite{BMPTY,HKT,BGZ}. They can also realize $5d$ versions of $A$ type class S theories\cite{BB}, and thus can be used to study them, and there are many other applications.

The purpose of this article is to study brane webs in the presence of an orientifold $5$-plane. First, this allows constructing $SO(N)$ and $USp(2N)$ gauge theories with fundamental matter, as first done in \cite{KB}. This can then be used to study these gauge theories. These systems can also be realized using an orientifold $7$-planes, as done in \cite{BZ1}, and our results agree with their finds.

More interestingly we can use this to realize more elaborate theories. First, we can realize a linear quiver of alternating $SO$ and $USp$ groups connected by half-bifundamentals. This then provides evidence that these theories exist as fixed points, and allows us to study some of their properties. Second, a subset of these theories are closely related to quivers of $SU$ group in the shape of the Dynkin diagram of type $D$ (henceforward referred to as $D$ shaped quiver), so these methods can be used to study these theories as well. We also argue that these can even be used to engineer $SO(N)$ gauge theories with spinor matter for $N\leq 12$. 


The structure of this article is as follows. Section 2 introduces the general construction in the more simplified case realizing a single gauge group. In section 3 we consider the general case giving a linear quiver with alternating $SO$ and $USp$ groups. We also consider the S-dual system leading to a $D$ shaped quiver of $SU$ gauge groups. Section 4 deals with describing $SO$ gauge groups with spinor matter. We end with some conclusions. The appendix provides a short review of index calculations and instanton counting.

\section{The general construction}

The starting point is the ordinary brane webs used to describe $\mathcal{N}$$=1$ supersymmetric $SU$ groups and their quivers \cite{HA,AHK}. The supersymmetry permits adding an $O5$-plane parallel to the D5-branes. This should lead to orthogonal or sympletic groups and their quivers (such a system was previously considered in \cite{KB}). Specifically, the construction involves an $O5$-plane with several parallel D5-brane crossed by NS5-branes. The orientifold enforces an orbifolding on the transverse coordinates which must be respected by the web. 


Next, we wish to recall several properties of $O5$-planes that will play an important role in what follows. There are $4$ different variants of $O5$-planes denoted as $O5^+, O5^-, \tilde{O5}^-$ and $\tilde{O5}^+$ \cite{HK}. Putting $N$ D5-branes on top of an $O5^+, \tilde{O5}^+$ results in a $USp(2N)$ gauge theory on them while putting them on top of an $O5^-$ results in an $SO(2N)$ gauge group. The $\tilde{O5}^-$ is an $O5^-$ plane with a stuck D5-brane and so putting $N$ D5-branes on top of it results in an $SO(2N+1)$ group. The $O5$-planes carry D5-brane charge: the $O5^+$ and $\tilde{O5}^+$ carry charge $+1$, the $O5^-$ carries charge $-1$ and the $\tilde{O5}^-$ carries charge $-\frac{1}{2}$. When a stuck NS5-brane crosses the $O5$-plane, in a way preserving $\mathcal{N}$$=1$ supersymmetry, it partitions the $O5$-plane into two parts of differing types: $O5^+ (\tilde{O5}^+)$ changes to $O5^- (\tilde{O5}^-)$. Likewise a stuck D7-brane also has a similar effect, now changing an $O5^+ (O5^-)$ into an $\tilde{O5}^+ (\tilde{O5}^-)$ and vice versa.

\begin{figure}
\center
\includegraphics[width=0.8\textwidth]{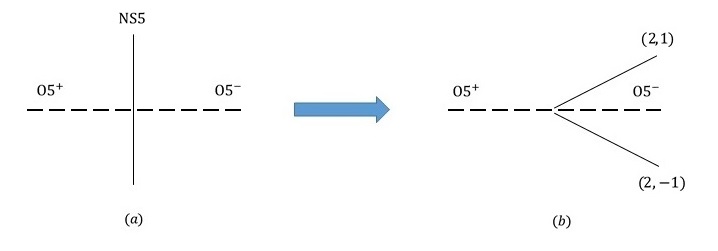} 
\caption{(a) The classical picture of a half NS5-brane crossing an $O5$-plane, represented as a black dashed line. (b) The quantum picture, where bending occurs so that the D5-brane charge is conserved.}
\label{fgr2}
\end{figure}

The change in the type of $O5$-plane when crossing an NS5-brane has important implications once quantum effects are taken into account. For example, let's consider a system consisting of an $O5^+$ with a stuck NS5-brane that changes it to an $O5^-$. Because of the type change, there is a jump in the D5-brane charge across the NS5-brane which should cause the NS5-brane to bend. Then taking into account charge conservation, sypersymmetry and invariance under the orbifolding, one concludes that the correct configuration should be a $(2,-1)$-brane crossing the $O5$-plane and becoming an $(2,1)$-brane. This is exhibited in figure \ref{fgr2}.

Next we explore the implications for the simplest cases of $N$ D5-branes stretched between two NS5-branes in the presence of one of the $O5$-plane types.

\subsection{$O5^+$ and USp groups}

\begin{figure}
\center
\includegraphics[width=1\textwidth]{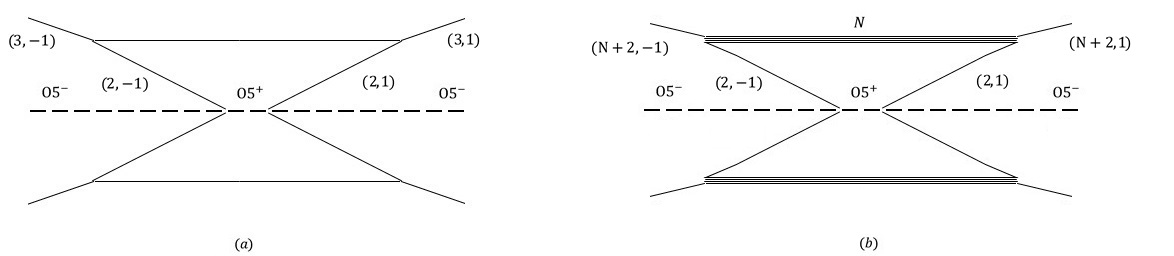} 
\caption{Brane webs for pure $USp(2N)$ gauge theory using an $O5^+$-plane. (a) Shows the $N=1$ case while (b) shows the general case.}
\label{fgr3}
\end{figure}

\begin{figure}
\center
\includegraphics[width=1\textwidth]{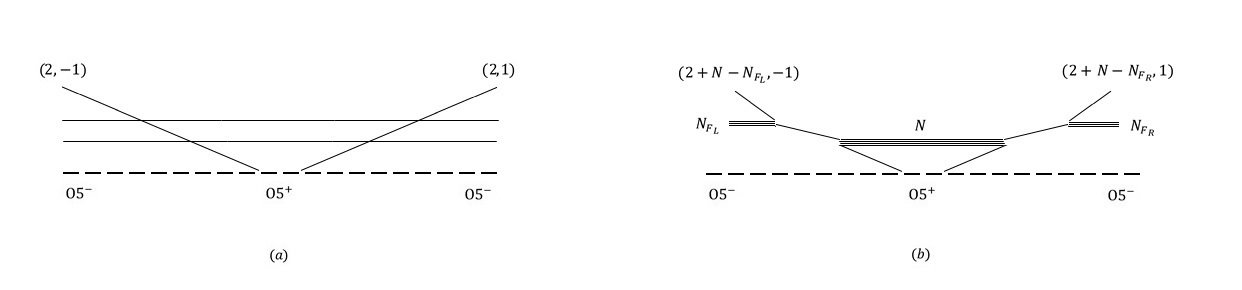} 
\caption{Examples of webs for a $USp$ gauge theory with fundamental flavor. (a) Shows the case of $USp(4)+4F$ while (b) shows the general case of $USp(2N)+(N_{F_L} + N_{F_R})F$.}
\label{fgr4}
\end{figure}

The classical picture consists of an $O5^+$-plane with $N$ D5-branes suspended between two stuck NS5-branes resulting in a $USp(2N)$ gauge theory. Taking into account the bending caused by the D5-branes and the $O5$-plane results in the web shown in figure \ref{fgr3}. It is now straightforward to generalize to cases with fundamental flavor by adding $(1,0)$ $7$-branes on top of the D5-branes. These can then be pulled through the external 5-branes, accompanied by Hanany-Witten transitions resulting in webs with semi-infinte external D5-branes. Examples of these are shown in figure \ref{fgr4}. Note that the gauge symmetry on $N_f$ external D5-branes is $SO(2N_f)$, since these sit on top of an $O5^-$, which is the correct global symmetry of $USp(2N)+N_fF$.

These webs can now be used to study a variety of issues in 5d gauge theories, notably, the existence of fixed point, identifying decoupled states in index calculations and motivating dualities. Such a thing was done for a different realization of this theory using $O7$-planes in \cite{BZ1}. One can see that the webs in the reduced space in these systems are similar to the ones, in the reduced space, with the $O5^+$-plane. Thus, most of the results found using the construction with the $O7$-plane are also true in this case, and we will not repeat them.

\begin{figure}
\center
\includegraphics[width=1\textwidth]{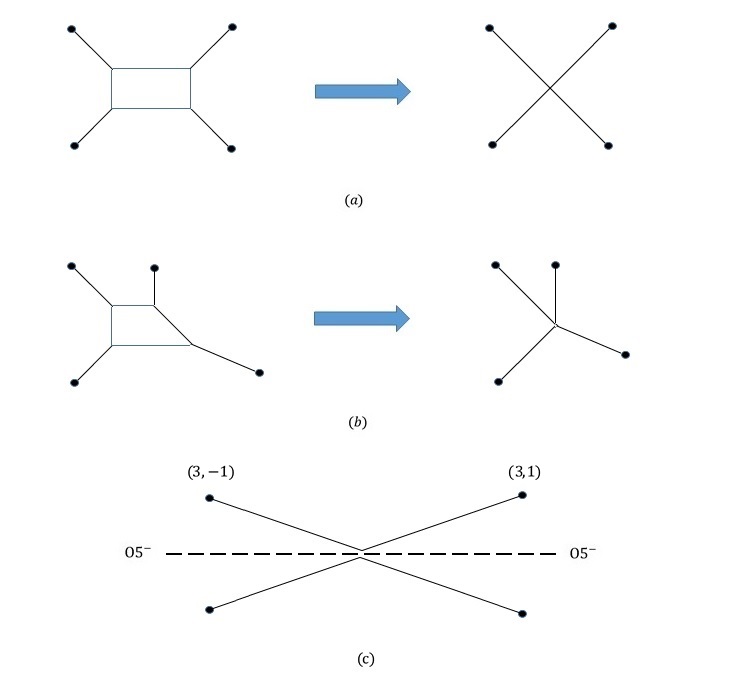} 
\caption{(a) The brane web for $SU_0(2)$, also known as the $E_1$ theory, at a generic coupling and at a generic point on the Coulomb branch (on the left), and at the origin of the Coulomb branch and taking the bare coupling to infinity (on the right), describing the fixed point. We have also explicitly drawn the $7$-branes, shown as black circles. These span the $8$ directions coming out of the picture. (b) The brane web for $SU_{\pi}(2)$, also known as the $\tilde{E}_1$ theory. The web on the left is for a generic coupling and at a generic point on the Coulomb branch, and the the web on the right is at the origin of the Coulomb branch and taking the bare coupling to infinity. One note that there is no Higgs branch in this case. (c) The web of figure \ref{fgr3} (a) at the origin of the Coulomb branch and infinite coupling constant.}
\label{fgr5}
\end{figure}

We do wish to discuss the manifestation of the Higgs branch in this web, as this is different from the $O7$-plane construction. In brane webs, the Higgs branch consists of all the possible motions of the 5-branes along the 7-branes. Uniquely for 5d, at the fixed point there can be additional Higgs branch directions besides the ones visible in the perturbative gauge theory\cite{CHFM}. Figure \ref{fgr5} (a) illustrates an example of this for the case of the $E_1$ theory: the fixed point has a one dimensional Higgs branch corresponding to separating the $(1,1)$ $5$-brane from the $(1,-1)$ $5$-brane. This theory can also be constructed using an $O5^+$-plane, as shown in figure \ref{fgr5} (c). This also exhibits the $1$-dimensional Higgs branch, now given by pulling the $(3,1)$ and $(3,-1)$ $5$-branes out of the $O5$-plane.

\begin{figure}
\center
\includegraphics[width=1.05\textwidth]{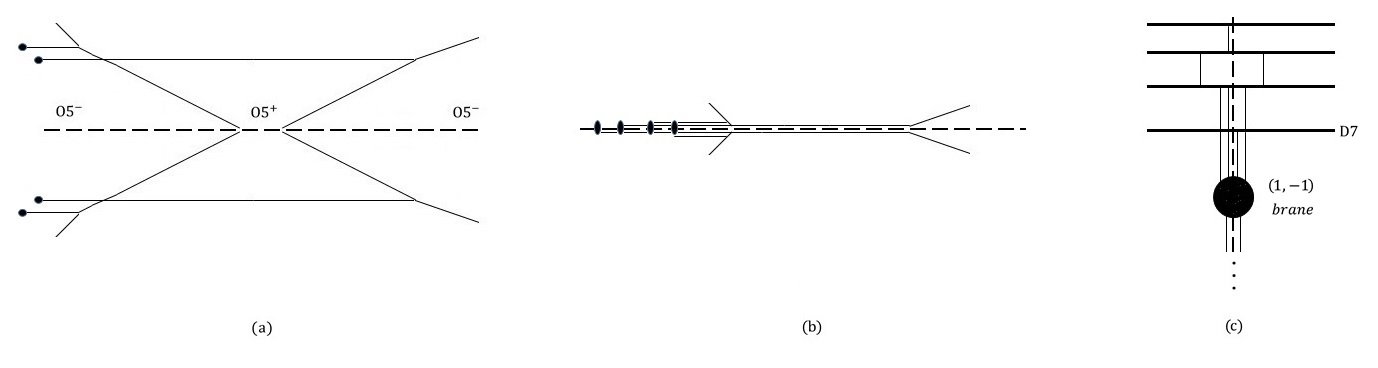} 
\caption{(a) The web for $USp(2)+2F$. (b) The web at the origin of the Coulomb branch and for massless flavor, after the D7-branes were separated along the $O5$-plane. (c) The web at a generic point on the Higgs branch. For ease of presentation we have used a different view of the web where the vertical straight lines are the D5-branes, the horizontal wide lines are the D7-branes, and the black dot is the (1,-1) 5-brane.}
\label{fgr7}
\end{figure}

The addition of fundamental flavor is done by adding $(1,0)$ $7$-branes, which we pull out, resulting in $N_f$ semi-infinte D5-branes all in the same direction, as shown in figure \ref{fgr7} (a). To enter the Higgs branch we go to the origin of the Coulomb branch and set the masses of the flavors to zero by coalescing all the D5-branes on the $O5$-plane. Furthermore, we separate the D7-branes along the $O5^-$-plane. When the $O5^-$-plane is crossed by a D7-brane it changes into an $\tilde{O5}^-$ which is an $O5^-$-plane with a stuck D5-brane. We thus conclude that upon each crossing one 5-brane must end on a D7-brane resulting in the picture shown in figure \ref{fgr7} (b). 

\begin{figure}
\center
\includegraphics[width=1\textwidth]{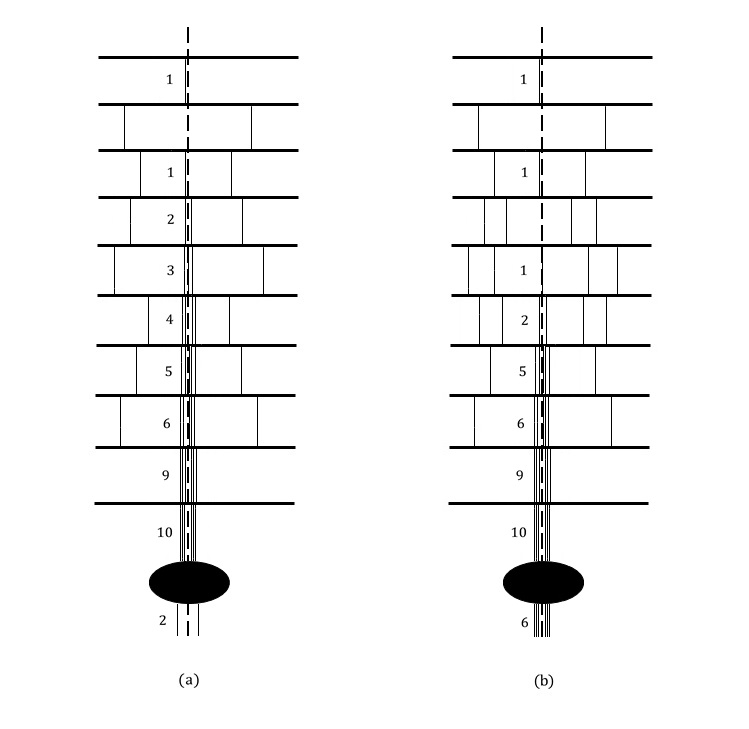} 
\caption{(a) The web at a generic point in the Higgs branch for $USp(2)+5F$. The numbers next to the $O5$-plane represent the number of D$5$-branes stuck on it. The Higgs branch is given by detaching a D$5$-branes and its image from the $O5$-plane. One can see that the Higgs branch is $7$ dimensional (quaternionic) in accordance with the gauge theory result. (b) The web at a generic point in the Higgs branch for $USp(6)+5F$. The Higgs branch for this theory is $10$ dimensional, again in accordance with the gauge theory result. Note that this differs from the case of (a) only by the number of color branes which is large enough so that every D7-brane can be connected to the other NS5-brane.}
\label{fgr8}
\end{figure}

The Higgs branch now consists of breaking the 5-branes on the 7-branes as shown in figure \ref{fgr7} (c). The possible breakings are limited by the S-rule, which necessitates that at most one 5-brane can be stretched between any given NS5-brane and D7-brane. When $N_f>2N$ this becomes more stringent as one can no longer connect a D7-brane to the other NS5-brane. Counting the possible breakings, with these restrictions, we indeed find the correct dimensions of the Higgs branch expected from the gauge theory. Additional examples of this are shown in figure \ref{fgr8}. 

Finally we take the fixed point limit by collapsing the gauge D5-branes. In this limit additional directions become available. First there is the $1$ dimension given by detaching the 5-branes from the $O5$-plane, similarly to the one in the $E_1$ theory. This exists for any number of flavors. When $N_f>2N$ there are further additional directions. It appears that when the NS5-branes touch, a D7-brane can always be connected to the other NS5-brane. This eases the constraints imposed by the  S-rule and allows additional directions. As we shall soon show this is necessary in order to recover the correct Higgs branch dimensions of known theories, like the rank $1$ $E_6$ theory. When $N_f=2N+4$ the two external NS5-branes become parallel and there is an additional direction given by breaking one of them on a $(0,1)$ 7-brane. Finally, when $N_f=2N+5$ there are intersecting external legs, where resolving the interaction leads to one of these external legs becoming a D$5$-brane due to passing thorough the monodromy of the other (see \cite{BZ1} for the details). At the fixed point, one can then also break this D$5$-brane on the $7$-branes leading to additional directions.

We can count the dimension of the Higgs branch for these theories and compare it with the one found using a different realization of these theories, for example using ordinary webs, finding complete agreement. As an example, consider $SU(2)$ with five flavors, the rank $1$ $E_6$ theory. As shown in figure \ref{fgr8} (a) the perturbative Higgs branch is $7$ dimensional, which is indeed the gauge theory result. An important limitation here is that there are only two gauge D$5$-branes so only two D$7$-branes can be connected to the other NS$5$-brane. This makes the constraint imposed by the S-rule more stringent. We have argued that this constraint should be relaxed at the fixed point, where the two NS$5$-branes coalesce. Indeed, without this constraint we would get a $10$ dimensional Higgs branch, as can be seen by comparing with the perturbative component of the Higgs branch for a different theory with the same number of flavors but with $2N>N_f$ like the $USp(6)+5F$ theory in figure \ref{fgr8} (b). Thus, the non-perturbative Higgs branch has $10+1=11$ (remember there is an additional direction given by taking the web off the $O5$-plane) which is indeed the Higgs branch dimension of the rank $1$ $E_6$ theory. Similarly, one also get from the web the correct dimensions of both the perturbative and non-perturbative Higgs branches for the $E_7$ and $E_8$ theories.     

\subsection{$O5^-$ and SO groups}

\begin{figure}
\center
\includegraphics[width=0.7\textwidth]{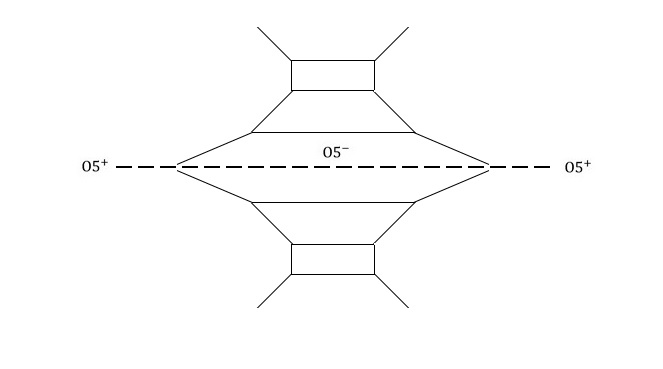} 
\caption{The web for a pure $SO$ gauge theory, in this case $SO(6)$, using an $O5^-$ plane.}
\label{fgr9}
\end{figure}

\begin{figure}
\center
\includegraphics[width=0.9\textwidth]{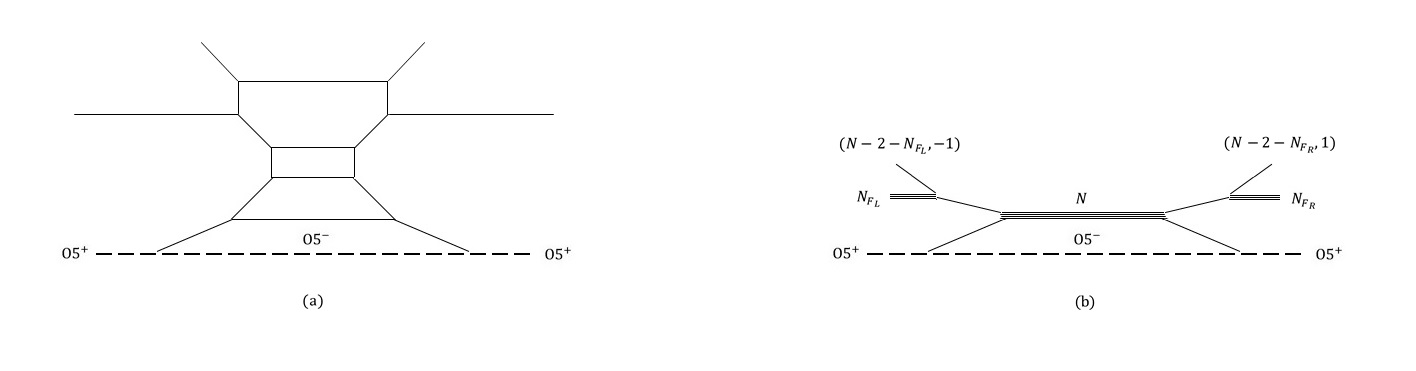} 
\caption{Examples of webs for an $SO$ gauge theory with fundamental flavor. (a) Shows the case of $SO(8)+2F$ while (b) shows the general case of $SO(2N)+(N_{F_L} + N_{F_R})F$.}
\label{fgr10}
\end{figure}

Changing the $O5^+$ to an $O5^-$ leads to an $SO(2N)$ gauge theory, an example of which is shown in figure \ref{fgr9}. The major difference in the web is that the bending caused by the $O5$-plane is in the opposite direction. This results in different bounds for fixed points, so for example one cannot draw a web for pure $SO(2)$, while one existed for the $O5^+$ case, pure $USp(2)$. This is in accordance with the expected UV incompleteness of 5d $U(1)$ gauge theories. The generalization by the addition of fundamental (in the vector representation) flavors is straightforward, examples shown in figure \ref{fgr10}. Now, the flavor D5-branes sit on top of an $O5^+$-plane resulting in a $USp(2N_f)$ global symmetry again in accordance with the gauge theory expectation.

An alternative realization of $SO(2N)+N_fF$ using an $O7$-plane also exists, and again the reduced space webs of these two constructions are similar. Thus, all the results seen from that construction are also valid in this one and we will not repeat them here. We do wish to describe the manifestation of the Higgs branch in this case. In the pure case, exactly as in the $O5^+$ case, one finds a 1-dimensional Higgs branch that opens at the fixed point. This agrees with the results seen also from the $O7$-plane constructions as well as other constructions when these are available (such as pure $SO(6)=SU(4)$). 

\begin{figure}
\center
\includegraphics[width=1.05\textwidth]{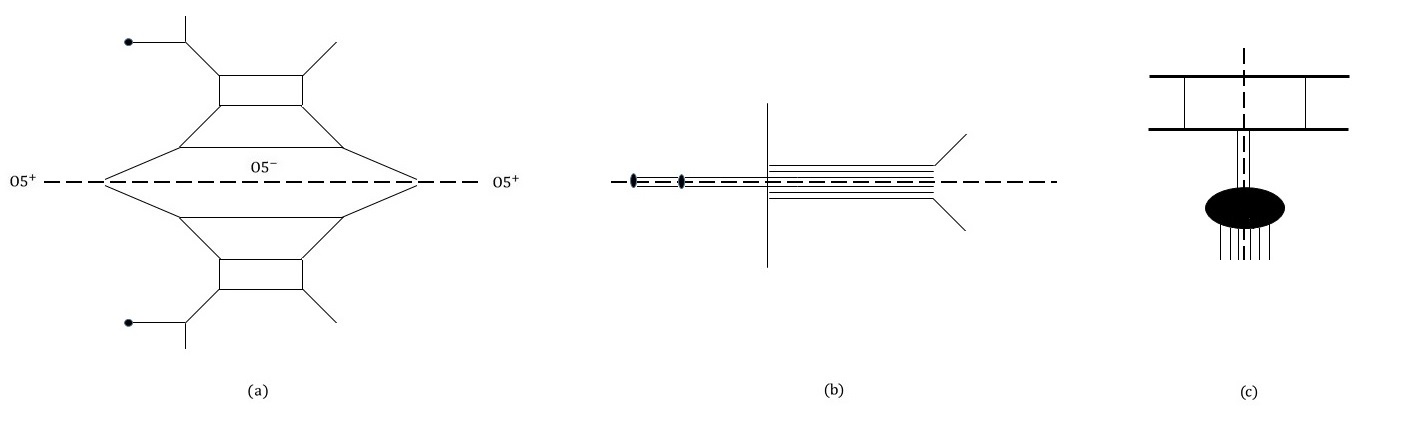} 
\caption{(a) The web for $SO(6)+1F$. (b) The web at the origin of the Coulomb branch and massless flavor, after separating the 7-branes. Since a D5-brane cannot be stuck on an $\tilde{O5}^+$ we are forced to stretch an extra D5-brane between the two 7-branes. (c) The web at a generic point on the Higgs branch.}
\label{fgr11}
\end{figure}

\begin{figure}
\center
\includegraphics[width=1.05\textwidth]{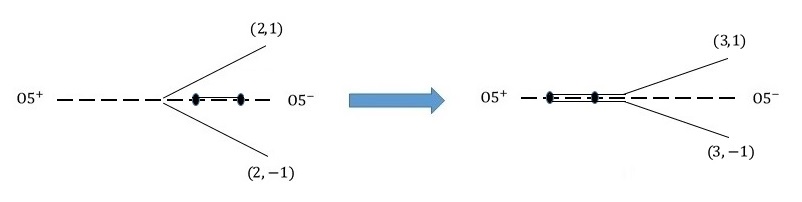} 
\caption{Separating a D$7$-brane and its image across an $O5^-$-plane, and then moving them past the $(2,1)$ $5$-brane. This leads to the configuration identical to separating a D$7$-brane and its image, each with a D$5$-brane ending on it, across an $O5^+$-plane.}
\label{XX10}
\end{figure}

Next we discuss the generalization when flavors are present, starting with the case of one flavor. We can again go to the origin of the Coulomb branch and the limit of zero mass. The major difference from the $USp$ case is encountered when separating a 7-brane from it's mirror image, since now there is an $\tilde{O5}^+$ between them. One cannot have a stuck D5-brane on an $\tilde{O5}^+$, so we conclude that when separating the 7-branes the two D5-branes must end on the same 7-brane\footnote{One can also arrive to the same conclusion by moving the D5-branes past the NS5-brane, and separate them on the $O5^-$ as done in the previous section. The separated 7-branes can now be moved back past the NS5-brane, along the $O5$-plane, resulting in the same outcome (see figure \ref{XX10}).}. The resulting construction is shown in figure \ref{fgr11}. The implication of this is that now there is a 1-dimensional Higgs branch where the gauge theory is broken to $SO(2N-1)$, in accordance with the gauge theory expectation. 

The generalization to more than one flavor is now straightforward. There are now $2N_f$ 7-branes stuck on the $O5$-plane which come in alternating pairs with one with no 5-branes ending on it and the other with two 5-branes ending on it. Breaking the 5-branes on the 7-branes, while taking due care of the S-rule, correctly reproduces the breaking pattern and the dimension of the Higgs branch as expected from the gauge theory. 

Finally, we take the fixed point limit and consider non-perturbative Higgs branch directions. Like in the $USp$ case, there is always the direction given by separating the external branes. However, in this case there does not appear to be a web with $N_f>2N-3$ so the directions associated with this case do not arise. When $N_f=2N-4$ the external legs become parallel while for $N_f=2N-3$ the external legs becomes intersecting, and there are extra directions similar to the cases of $N_f=2N+4, 2N+5$ in the $USp$ case.

\subsection{$\tilde{O5}$}

\begin{figure}
\center
\includegraphics[width=1.05\textwidth]{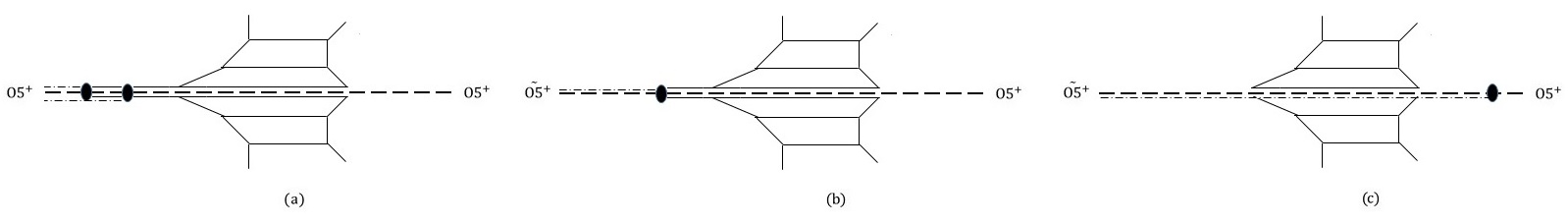} 
\caption{(a) The web for $SO(6)+1F$ where we have also drawn the monodromy line of the $7$-branes, depicted as a thin dot-dash line. (b) The web for $SO(5)$ which one gets by going on the Higgs branch of $SO(6)+1F$. (c) Pulling the 7-brane to the other side through two HW transitions results in the web of $SO(5)$ using an $\tilde{O5}^-$-plane. Note that there is an half monodromy stuck on the $\tilde{O5}^-$-plane which fixes the D5 brane charge conservation. When drawing an $\tilde{O5}^-$ plane we also draw the stuck D$5$-branes, slightly above the $\tilde{O5}^-$ plane, and the half monodromy line, slightly below the $\tilde{O5}^-$-plane.}
\label{fgr13}
\end{figure}

Finally we want to consider the case of an $\tilde{O5}$ plane. At first one encounters a problem with the fractional NS5-brane on it. This leads to a change between an $\tilde{O5}^+$ and an $\tilde{O5}^-$, resulting in a jump in the D5-brane charge. However this jump is now fractional, and it does not appear to be possible to reconcile the bending required by charge conservation with the reflection symmetry implied by the orientifold. One approach to realizing these theories, in the case of an $\tilde{O5}^-$, is to use the construction for $SO(2N+2)+1F$ and go on the Higgs branch leading to $SO(2N+1)$. An example of this is shown in figure \ref{fgr13}. We can now transform to a description with an $\tilde{O5}^-$ by moving the stuck D7-brane though the external 5-branes to the other end of the $O5$-plane, while taking due care of the monodromy of the 7-brane. Now, the resolution of the above issue is clear: there is also a half monodromy on the $O5$-plane which corrects the bending so as to be consistent with the orbifolding.

One can now use this construction to realize $SO(2N+1)+N_fF$ gauge theories. Once again the reduced space web matches the one using an $O7$-plane and the results found from this description also apply to this case. The Higgs branch is realized exactly as in the $SO(2N)$ case.

\begin{figure}
\center
\includegraphics[width=1.05\textwidth]{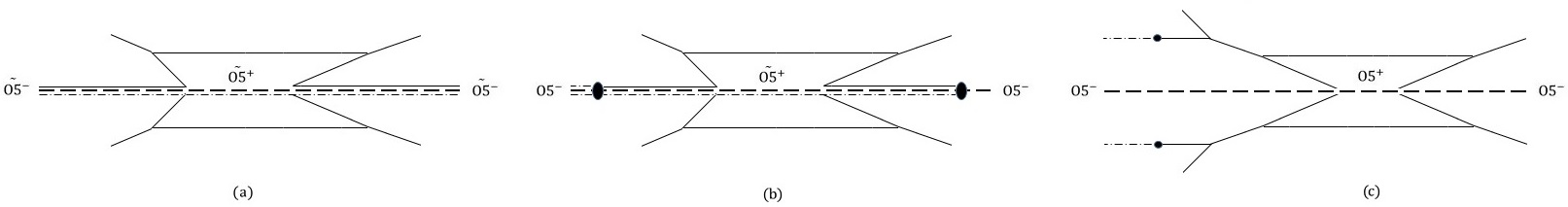} 
\caption{(a) The web for $USp(2)$ using an $\tilde{O5}^+$-plane. Like with the $\tilde{O5}^-$-plane, we also draw the half monodromy line slightly below the $\tilde{O5}^+$-plane. (b) The same web where we have added the half 7-branes on which the half D5-branes end. (c) The same web after taking the 7-branes past the NS5-brane, merging them to a full 7-brane and pulling it out of the $O5$ plane. One can see that this describes $USp(2)+1F$.}
\label{fgr14}
\end{figure}

Finally, we can inquire about the web with an $\tilde{O5}^+$. A natural guess is that this describes the $USp$ theory with $\theta=\pi$ \footnote{We think $O5^+$ describes the $\theta=0$ case based on the Higgs branch. As mentioned in figure \ref{fgr5}, the $E_1$ fixed point has a 1-dimensional Higgs branch while $\tilde{E}_1$ does not.}. Indeed, we expect such theories to exist yet there are no discrete choices in the web save for this. Furthermore, such a thing occurs for example in the construction of the maximally supersymmetric 5d $USp(2N)$ gauge theory using $O4$-planes\cite{Hori,Tachi}. However, the web for $USp(2N)$ using an $\tilde{O5}^+$, an example of which is shown in figure \ref{fgr14}, appears to be identical to $USp(2N)+1F$ and does not describe a new theory. Thus, it appears that there is no difference between using an $O5^+$ and an $\tilde{O5}^+$. 

As a result, there does not appear to be a way to describe $USp_{\pi}(2N)$ with an $O5$-plane. One can set out to get such a web by starting with the $E_2$ web and giving a mass to a flavor. Then depending on the mass sign one gets either the $E_1$ or $\tilde{E}_1$ theories. This indeed works for ordinary webs and ones with an $O7$-plane, but not for this case. The problem is that in this case the flavor can only be integrated in one direction. It is interesting whether there is a fundamental reason why $\theta=\pi$ cannot be accommodated in this construction, or alternatively if it is possible to incorporate it in the web in a more intricate way.  

\section{$SO \times USp$ quivers}

\begin{figure}
\center
\includegraphics[width=0.9\textwidth]{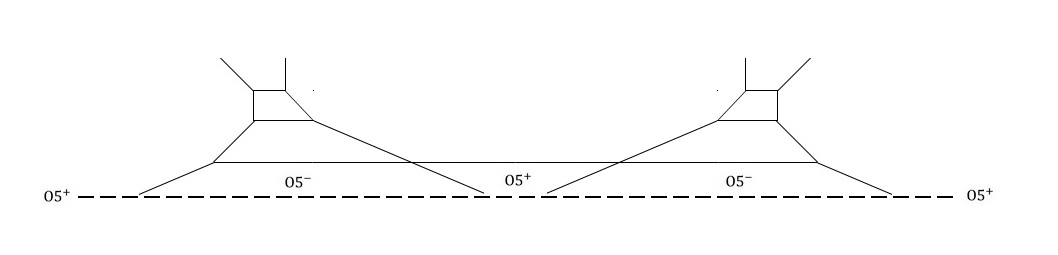} 
\caption{The web for $SO(6)\times USp(2) \times SO(6)$ where a half-bifundamental is understood to exist whenever an $\times$ is used between an $SO$ and $USp$ groups.}
\label{fgr15}
\end{figure}

\begin{figure}
\center
\includegraphics[width=0.9\textwidth]{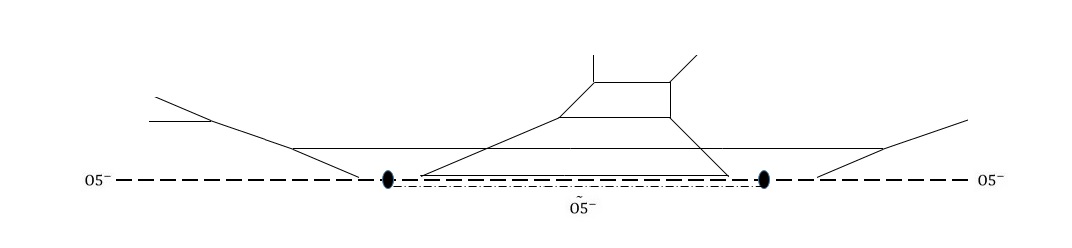} 
\caption{The web for $3HF+USp(2)\times SO(7) \times USp(2)+1HF$ where $HF$ stands for a half-fundamental. Note that the total number of half-fundamentals is even for each $USp(2)$ so there is no gauge anomaly.}
\label{fgr16}
\end{figure}

So far we considered a system with just two external NS5-branes. The web can be generalized to an arbitrary number of such branes which leads to a long quiver with alternating $SO$ and $USp$ groups connected by half-hypers in the bifundamental representation. Two examples of this are shown in figures \ref{fgr15} and \ref{fgr16}. The quiver can contain both $SO(2N)$ and $SO(2N+1)$ gauge groups which can be achieved by adding a stuck D7-brane. This can also lead to an half-hyper in the fundamental for $USp(2N)$ which appears in the web as a stuck D7-brane or a stuck external D5-brane. Note that due to a global gauge anomaly, the 5d version of \cite{Wit}, a $USp(2N)$ gauge theory must have an even number of fundamental half-hypers\cite{SMI}. This is indeed respected by the web. 


These webs have several interesting implications. First, they point to the existence of fixed point theories for these quiver theories. This can also be generalized by adding flavors and the web can be used to argue the limit beyond which a fixed point does not exist. The web can also be used to study the Higgs branch, both the perturbative component and the non-perturbative component, as in the previous examples. It can also be used to identify decoupled states in index calculations using the $SO \times USp$ formalism, as done for other systems in \cite{BMPTY,HKT,BGZ,BZ1}. 


Yet another application is to study symmetry enhancement in such theories. In brane webs this is manifested by some of the external legs becoming parallel. For example, consider a linear quiver consisting of $n_G$ groups of ranks $N_i$, for $i=1,2...,n_G$, with $F_i$ fundamental hypers under the $i$'th group. By inspecting the D$5$-charges of the NS$5$-charge carrying external branes, one can see that these are neither parallel nor intersecting as long as $2N_i\pm 4 > F_i + \frac{N_{i-1}+N_{i+1}}{2}$ where the $+$ sign is for $USp$ groups and the $-$ sign is for $SO$ groups, and we take $N_0 = N_{n_G+1} = 0$. 

\begin{figure}
\center
\includegraphics[width=0.9\textwidth]{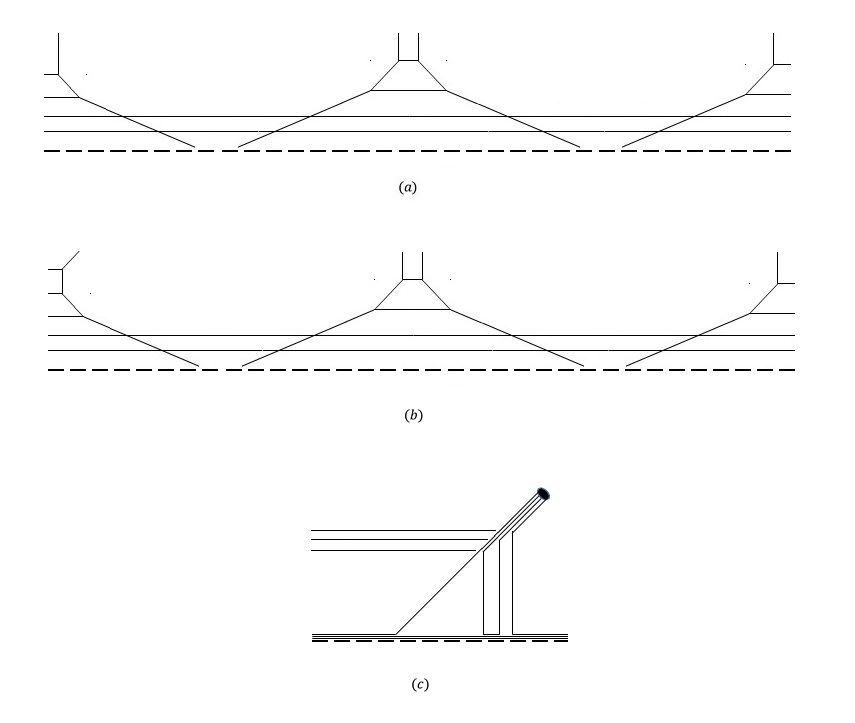} 
\caption{(a) The web for $4F+USp(4)\times SO(8) \times USp(4)+4F$ where each group obeys $2N_i\pm 4 = F_i + \frac{N_{i-1}+N_{i+1}}{2}$. One can see that the web as $4$ parallel NS$5$-branes suggesting an enhancement of $U_T(1)^3\rightarrow SU(4)$. (b) The web for $5F+USp(4)\times SO(8) \times USp(4)+4F$. One can see that the $(1,1)$ $5$-brane intersects the $3$ NS$5$-branes. Resolving the intersection via a Hanany-Witten transition leads to the web in (c). One can see that due to passing through the monodromy of the $(1,1)$ $5$-brane, the NS$5$-branes become D$5$-branes. This suggests that this fixed point as an enhanced $SO(16)$ global symmetry.}
\label{XX11}
\end{figure}

If however $2N_i\pm 4 = F_i + \frac{N_{i-1}+N_{i+1}}{2}$ for some series of adjacent groups then a group of NS$5$-charge carrying external branes become parallel, signaling an enhancement of the topological symmetries associated with these groups. In particular, if all the groups obey $2N_i\pm 4 = F_i + \frac{N_{i-1}+N_{i+1}}{2}$ then all these external branes become parallel, suggesting an enhancement of $U(1)^{n_G}\rightarrow SU(n_G+1)$ (see figure \ref{XX11} (a) for an example). 


If $2N_i\pm 4 < F_i + \frac{N_{i-1}+N_{i+1}}{2}$ occures for one of the groups then the NS$5$-charge carrying external branes associated with that group intersect (see figure \ref{XX11} (b) for an example). The intersecting branes can be continued past one another accompanied with a Hanany-Witten transition. If this process terminates after a finite number of such transitions then this $5d$ gauge theory go to a $5d$ fixed point described by the collapsed web (see figure \ref{XX11} (c) for an example). One can now use this brane web to try and read the global symmetry of the fixed point theory. 

For example, consider the previously considered class of theories, where every group obeys $2N_i\pm 4 = F_i + \frac{N_{i-1}+N_{i+1}}{2}$. Say we now add one more flavor to one of the edge groups, say for $i=1$. As seen in figure \ref{XX11} (b), this leads to a configuration with the leftmost NS$5$-charge carrying external brane intersecting the $n_G$ other NS$5$-charge carrying branes. This can be resolved by continuing them past one another, accompanied with a Hanany-Witten transition, leading to a configuration with no intersection (see figure \ref{XX11} (c) for an example) so we conclude that this class of theories go to a $5d$ fixed point. Inspecting the web one sees that this configuration as $F_1+n_G$ D$5$-branes on the left side and $F_{n_G}$  D$5$-branes on the right side. Thus, we conclude that in this case there should be an enhancement of $U(1)^{n_G}\times SO(2F_1)\rightarrow SO(2F_1+2n_G)$ or $U(1)^{n_G}\times USp(2F_1)\rightarrow USp(2F_1+2n_G)$ depending on whether the $i=1$ group is of type $USp$ or $SO$ respectively.

\begin{figure}
\center
\includegraphics[width=0.85\textwidth]{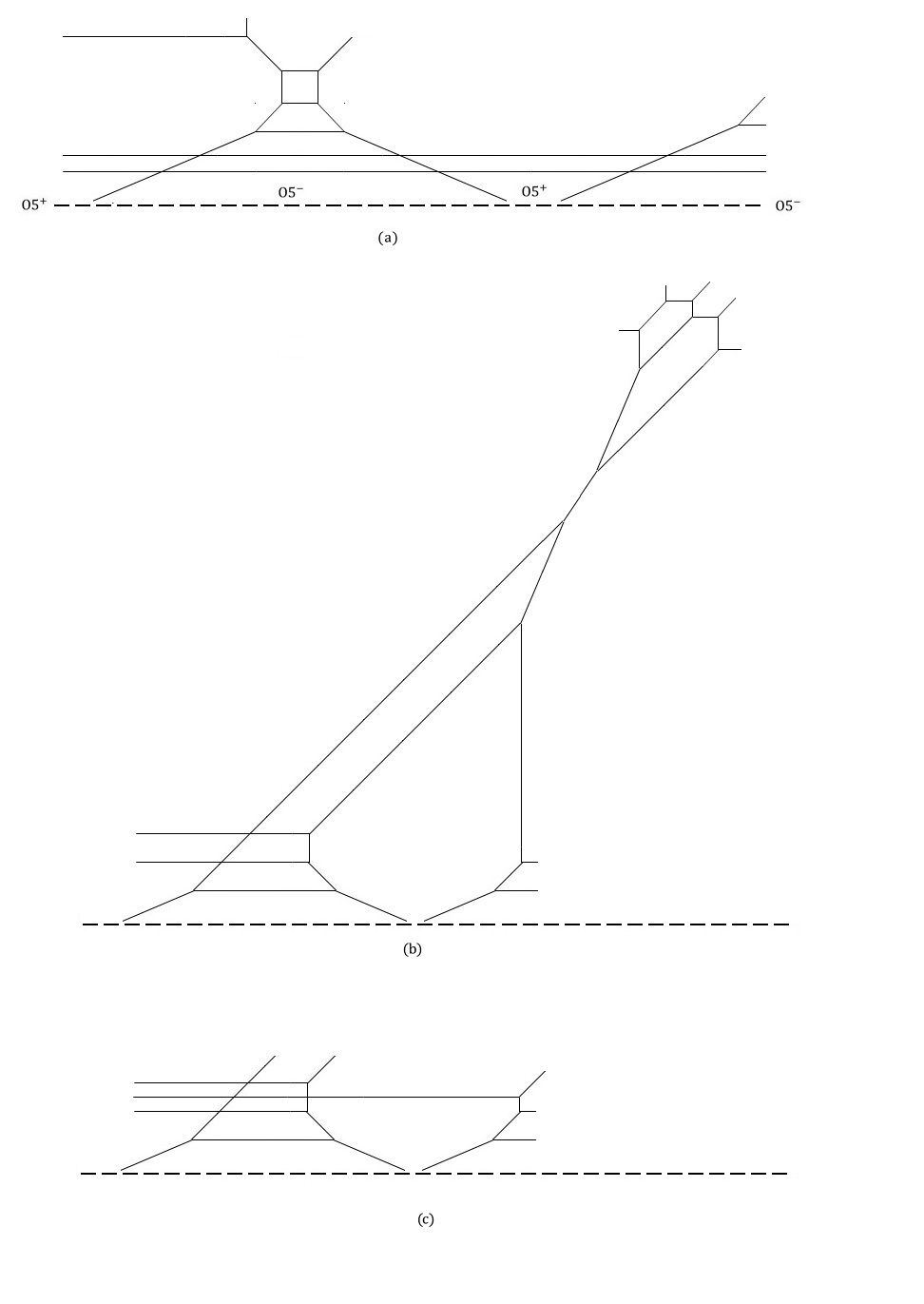} 
\caption{(a) The web describing an $3F+SO(10)\times USp(4)+3F$ gauge theory. (b) The same web after several flop transitions. One can see that it is identical to an $SU(3)+2F$ gauging the enhanced $SU(3)$ instantonic symmetry of the web in (c). (c) The web describing an $3F+SO(8)\times USp(2)+2F$ gauge theory.}
\label{fgr18}
\end{figure}

\begin{figure}
\center
\includegraphics[width=1.05\textwidth]{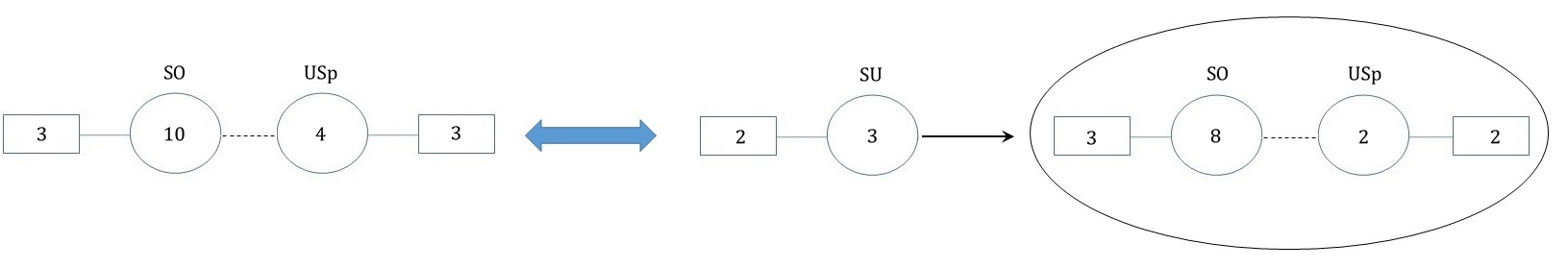} 
\caption{The duality implied by the manipulations of figure \ref{fgr18} where we use a dotted line for a half-hyper in the bifundamental representation. The gauged $SU(3)$ group on the right side is instantonic and so is not perturbatively visible.}
\label{fgr17}
\end{figure}

In the rest of this section we concentrate on one further application which is motivating 5d dualities. These are manifested by a different gauge theory description, of the same web, generically in a different $SL(2,Z)$ frame. We can first start by generalizing the dualities of \cite{BZ1} also to the quiver case. As a simple example consider the web shown in figure \ref{fgr18} (a) which describes a $3F+SO(10)\times USp(4)+3F$ gauge theory. The web can be deformed through several flop transitions to the one shown in figure \ref{fgr18} (b). The web shows an $SU(3)+2F$ gauge theory, existing on the $(1,1)$-branes, gauging a global $SU(3)$ symmetry in the remaining web shown in figure \ref{fgr18} (c), which describes a $3F+SO(8)\times USp(2)+2F$ gauge theory. The web suggests that this theory has an enhanced $SU(3)$ instantonic global symmetry. This leads to the duality shown in figure \ref{fgr17}. Since the gauged global symmetry is instantonic, this does not lead to a gauge theory duality. For this we would need to find a dual description of the theory in figure \ref{fgr18} (c), in which the $SU(3)$ global symmetry is perturbativly realized. We can consider generalizations of this duality, but in all we encounter the same problem, where we do not find a complete gauge theory description of the dual side.

\subsection{S-duality and $O5^0$-plane}

In order to make further progress one must do an $SL(2,Z)$ transformation on the entire system, including the $O5$-plane. This requires understanding the behavior of the $O5$-plane under these transformations, in particular S-duality. There is one case where this is actually known which is an $O5^{-}$-plane with a full D5-brane. In that case the total D5-brane charge of the system is zero, and its strong coupling behavior is of the perturbative orbifold $R^4/Z_2$, the $Z_2$ being a reflection in the four directions combined with $(-1)^{F_L}$\cite{HK,HZ}. Thus, in this case we might be able to say something explicit about the S-dual theory.

\begin{figure}
\center
\includegraphics[width=0.8\textwidth]{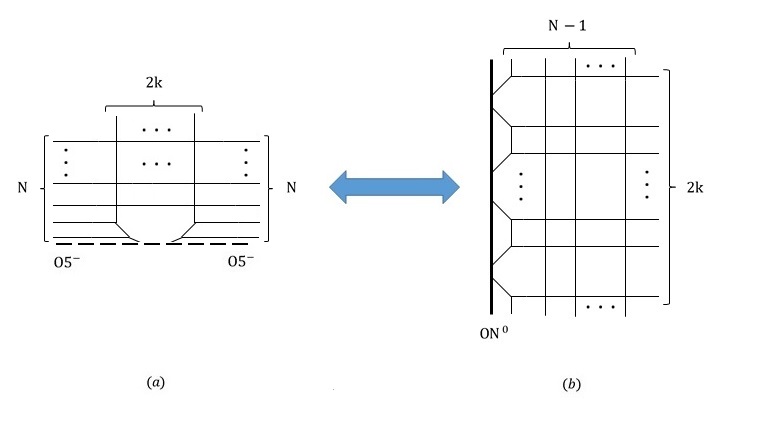} 
\caption{The brane configuration consisting of an $O5^-$ plane, $2k$ NS5-branes and $N$ D5-branes. (b) The S-dual of (a).}
\label{fgr22}
\end{figure}

\begin{figure}
\center
\includegraphics[width=0.8\textwidth]{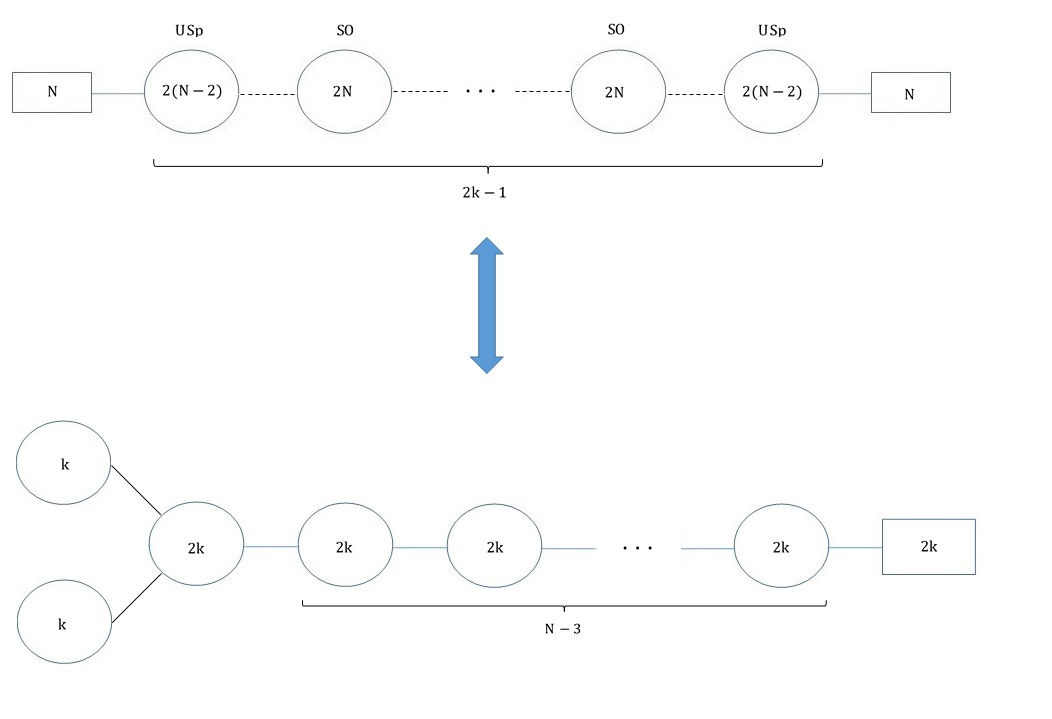} 
\caption{The duality implied from figure \ref{fgr22}. The upper gauge theory is the one described by \ref{fgr22} (a) while the lower one is the one described by \ref{fgr22} (b) (all the groups are $SU$ with CS level $0$).}
\label{fgr23}
\end{figure}

The simplest thing to start with is an $O5^-$-plane with $N$ parallel D5-branes crossed by $2k$ NS5-branes, where an even number is necessary so that asymptotically we still have an $O5^-$-plane so that we can apply S-duality. Doing S-duality on this configuration results in the web shown in figure \ref{fgr22} (b). This clearly shows a long $SU(2k)$ linear quiver where at its end there are the branes connecting the NS5-brane with the orbifold fixed plane. The gauge theory existing on such a system is known to be an $SU(n_1)\times SU(n_2)$ gauge theory with $n_1+n_2=2k$. The numbers $n_1, n_2$ arise as there are two different types of $5$-branes ending on the orbifold fixed plane differing by their charge under the twisted field living at the fixed plane. In the present case, one can see an $SU(k) \times SU(k)$ gauge theory. The matter content supplied by the crossed NS5-brane is a bifundamental hyper between each $SU(k)$ and the adjacent $SU(2k)$ so we conclude the duality shown in figure \ref{fgr23}. The Chern-Simons levels are all $0$ as the web is invariant under reflection which is the brane web analogue of charge conjugation in the gauge theory.

One can now inquire whether we can find additional evidence for this duality. First, we can do several simple checks such as comparing the dimension of the Coulomb branch and the number of mass parameters. A short counting shows that they are equal. Matching the global symmetries is harder as not all are classically apparent. The theory with the $O5^-$ plane classically has an $U(1)^{2k-1}\times SO(2N)^2$ which, as suggested by the web, is enhanced to $SO(2N)^2 \times SU(2k)$. However, on the dual side, the classical global symmetry is $U(1)^{2N}\times SU(2k)$, which the duality suggests should enhance to $SO(2N)^2 \times SU(2k)$. Indeed this is supported by the instanton analysis of \cite{Tachi1.5,Yon}.  

Both theories also have discrete global symmetries. The $D$-shaped quiver theory has a $Z_2 \times Z_2$ symmetry given by charge conjugation and exchanging the two $SU(k)$ groups. The $SO \times USp$ quiver theory as a $Z_2$ reflection symmetry, which the web suggests matches charge conjugation in the dual theory. The duality suggests that there should be another $Z_2$ that matches the exchange of the two $SU(k)$ groups. There is indeed an extra $Z_2$ on the $SO \times USp$ quiver theory, exchanging the two spinor representations of all the $SO$ groups. We expect this to match the exchange of the two $SU(k)$ groups. Note that this operation on the Dynkin diagram of type $D$ indeed corresponds to exchanging the two spinor representations. This matches similar results in the linear $A$ type quiver\cite{BGZ}.

\begin{figure}
\center
\includegraphics[width=0.8\textwidth]{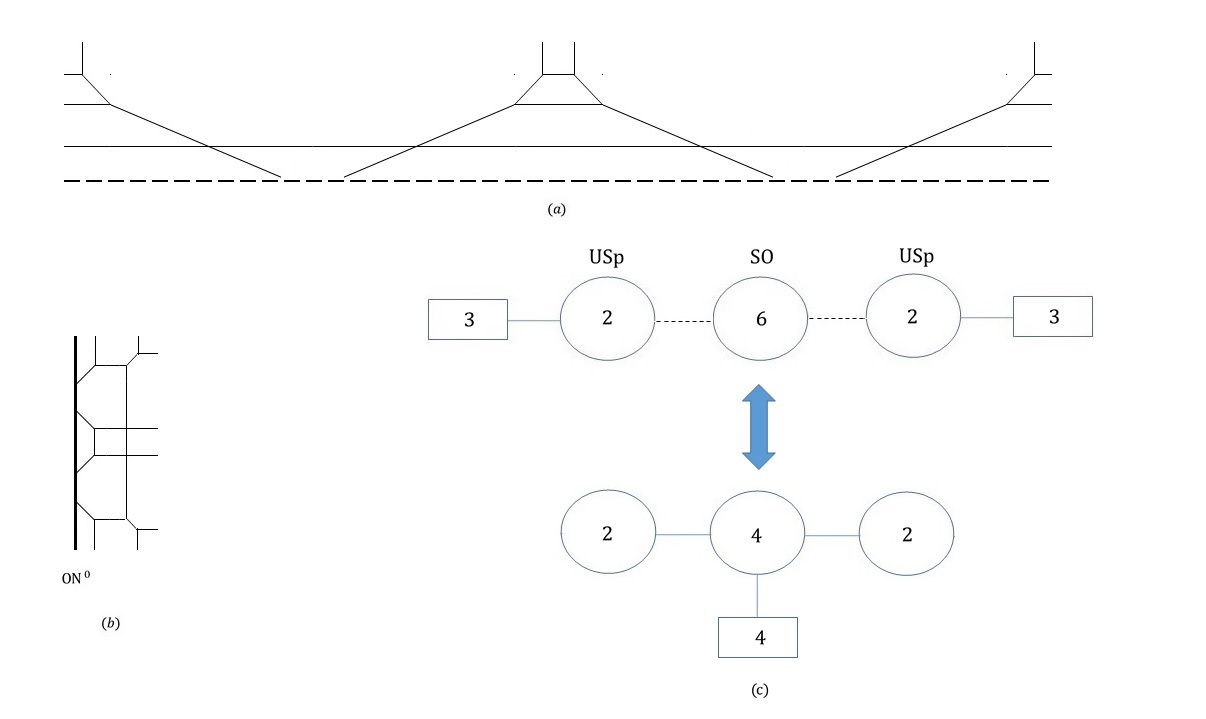} 
\caption{The duality in the case of $N=3, k=2$. (a) The web describing $3F+USp(2)\times SO(6) \times USp(2)+3F$. (b) The S-dual web describing the gauge theory $SU(2)\times SU(4) \times SU(2)$ with $4$ flavors for the $SU(4)$. (c) The resulting duality.}
\label{fgr24}
\end{figure}

Another check that can be done on this duality is comparing their superconformal indices (see the appendix for a review of the $5d$ superconformal index). As a starting point we can take the simplest example illustrated in figure \ref{fgr24}. On one side we have the gauge theory $SU(2)\times SU(4) \times SU(2)$ with $4$ flavors for the $SU(4)$. The Chern-Simons levels of both $SU(2)$ groups should be $0$ which corresponds to $\theta=0$. The web suggests this theory as an $SU(4)^3$ global symmetry where one is visible perturbativly and the others are brought by instantonic enhancement. Indeed by an index calculation we verified that such an enhancement is present. 

In the dual theory we also expect an $SU(4)^3$ global symmetry where now an $SO(6)^2 = SU(4)^2$ is perturbativly visible while instantons should provide the conserved currents for the third $SU(4)$. The web suggests that this requires the contributions of (2,0,0) + (0,1,0) + (0,0,2) + (2,1,0)+ (0,1,2) + (2,1,2) instantons. 

\begin{figure}
\center
\includegraphics[width=1\textwidth]{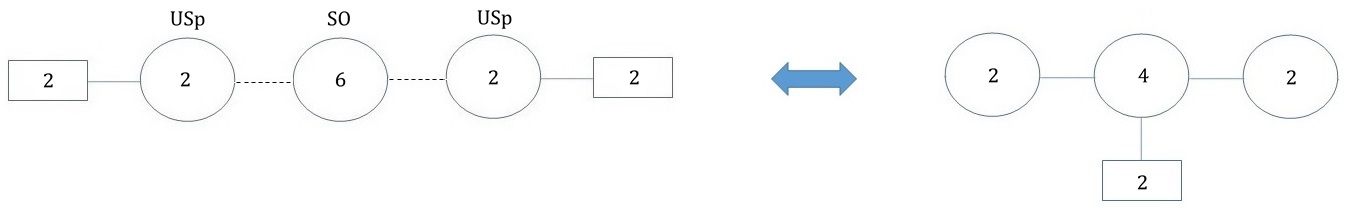} 
\caption{The duality one gets after integrating out two flavors from the duality of figure \ref{fgr24}.}
\label{fgr25}
\end{figure}

Unfortunately, calculating all of these contributions is technically demanding so it is worthwhile to look at a simpler example. Particularly, we can consider integrating out flavors so as to reduce the degree of enhancement. This is done by taking an external D5-branes to infinity. Note that for the two edge flavors this is identical to pulling an NS5-brane to infinity, and thus to also integrating out a flavor in the dual theory. Thus, we arrive at the duality shown in figure \ref{fgr25} which is the one we shall check using the superconformal index.

\begin{figure}
\center
\includegraphics[width=0.4\textwidth]{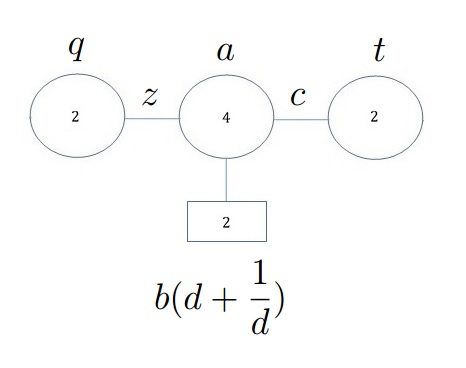} 
\caption{The fugacity spanning for the $SU(2)\times SU(4) \times SU(2)$ theory.}
\label{fgr26}
\end{figure}

The theory on the right of figure \ref{fgr25} is $SU(2)\times SU(4) \times SU(2)$ gauge theory with $2$ flavors for the $SU(4)$. The classical global symmetry is $SU(2) \times U(1)^6$. The fugacity allocation is shown in figure \ref{fgr26}. We work to order $x^5$ which requires the contributions of the (1,0,0) + (0,1,0) + (0,0,1) + (1,1,0) + (0,1,1) + (1,0,1) + (2,0,0) + (0,0,2) + (1,1,1) instantons. We find:

\bea
Index_{D-quiver} & = & 1 + x^2 \left( 7 + d^2 + \frac{1}{d^2} + (q + \frac{1}{q})(z^2 + \frac{1}{z^2}) + (t + \frac{1}{t})(c^2 + \frac{1}{c^2}) \right) \\ \nonumber & + &  x^3(y + \frac{1}{y})\left( 8 + d^2 + \frac{1}{d^2} + (q + \frac{1}{q})(z^2 + \frac{1}{z^2}) + (t + \frac{1}{t})(c^2 + \frac{1}{c^2}) \right) \\ \nonumber & + & O(x^4)
\eea
where we have only displayed terms up to order $x^3$ even though the calculation was done up to order $x^5$.
 
\begin{figure}
\center
\includegraphics[width=0.6\textwidth]{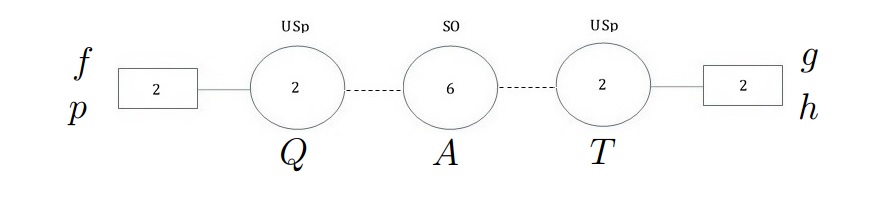} 
\caption{The fugacity spanning for the $USp(2)\times SO(6) \times USp(2)$ theory.}
\label{fgr27}
\end{figure}

From the $x^2$ terms in the index we see that the (1,0,0) + (0,0,1) instantons provide the conserved currents to enhance $U(1)^4\rightarrow SU(2)^4$ so that the quantum symmetry appears to be $SU(2)^5\times U(1)^2$. The remainder of the index indeed forms characters of the enhanced symmetry as will become apparent when we next compare it with the index of the $SO \times USp$ quiver. In that theory the classical global symmetry is $SU(2)^4\times U(1)^3$. We use the fugacity spanning shown in figure \ref{fgr27}. To the order we are working with we get contributions from the (1,0,0) + (0,1,0) + (0,0,1) + (1,0,1) + (0,2,0) + (1,1,1) instantons. We find:  

\bea
Index_{SO/USp} & = & 1 + x^2 \left( 7 + f^2 + \frac{1}{f^2} + g^2 + \frac{1}{g^2} + h^2 + \frac{1}{h^2} + p^2 + \frac{1}{p^2} + A + \frac{1}{A} \right) \\ \nonumber & + &  x^3(y + \frac{1}{y})\left( 8 + f^2 + \frac{1}{f^2} + g^2 + \frac{1}{g^2} + h^2 + \frac{1}{h^2} + p^2 + \frac{1}{p^2} + A + \frac{1}{A} \right) \\ \nonumber & + & O(x^4)
\eea
where again we displayed terms only up to $x^3$.

From the $x^2$ terms we see that the (0,1,0) instantons bring about an enhancement of $U(1)\rightarrow SU(2)$ so that the quantum symmetry is $SU(2)^5\times U(1)^2$ in accordance with the dual theory. Further comparing the two we find that taking: $\sqrt{A}=d$, $\frac{Q}{T} = b^2$, $Q T \sqrt{A} = a \sqrt{t q}$, $f=z\sqrt{q}$, $g=\frac{\sqrt{q}}{z}$, $h=\frac{\sqrt{t}}{c}$, $p=c\sqrt{t}$ render the two equal. From the matching we see that the enhanced $SU(2)$ global symmetry of the $SO \times USp$ quiver matches the perturbative $SU(2)$ global symmetry of the $D$ shaped quiver theory and vice versa, as expected from the web. 

The index also makes manifest the $Z_2 \times Z_2$ global symmetry. It acts on the theory by permutating the $4$ $SU(2)$ global symmetry groups, perturbativly realized in the $SO \times USp$ quiver, similar to its action on the $4$ vertices of a rectangle whose symmetry group is also $Z_2 \times Z_2$. The matching of fugacities is consistent with charge conjugation of the D-shaped quiver mapped to reflecting the $SO \times USp$ quiver, while exchanging the two $SU(2)$ groups is mapped to charge conjugation of the $SO \times USp$ quiver (which exchanges the two spinor representations of $SO(6)$).

The index can be written in characters of the $SU(2)^5\times U(1)^2$ global symmetry where it reads:

\bea
Index & = & 1 + x^2 (2+\chi[\bold{3},\bold{1},\bold{1},\bold{1},\bold{1}]+\chi[\bold{1},\bold{3},\bold{1},\bold{1},\bold{1}]+\chi[\bold{1},\bold{1},\bold{3},\bold{1},\bold{1}]+\chi[\bold{1},\bold{1},\bold{1},\bold{3},\bold{1}] \\ \nonumber & + &  \chi[\bold{1},\bold{1},\bold{1},\bold{1},\bold{3}]) + x^3 \chi_y[\bold{2}](3+\chi[\bold{3},\bold{1},\bold{1},\bold{1},\bold{1}]+\chi[\bold{1},\bold{3},\bold{1},\bold{1},\bold{1}] \\ \nonumber & + & \chi[\bold{1},\bold{1},\bold{3},\bold{1},\bold{1}]+\chi[\bold{1},\bold{1},\bold{1},\bold{3},\bold{1}]+\chi[\bold{1},\bold{1},\bold{1},\bold{1},\bold{3}]) \\ \nonumber & + & x^4 \left( \chi_y[\bold{3}](3+\chi[\bold{3},\bold{1},\bold{1},\bold{1},\bold{1}]+\chi[\bold{1},\bold{3},\bold{1},\bold{1},\bold{1}]+\chi[\bold{1},\bold{1},\bold{3},\bold{1},\bold{1}]+\chi[\bold{1},\bold{1},\bold{1},\bold{3},\bold{1}] \right. \\ \nonumber & + & \left. \chi[\bold{1},\bold{1},\bold{1},\bold{1},\bold{3}]) + \chi[\bold{5},\bold{1},\bold{1},\bold{1},\bold{1}]+\chi[\bold{1},\bold{5},\bold{1},\bold{1},\bold{1}]+\chi[\bold{1},\bold{1},\bold{5},\bold{1},\bold{1}]+\chi[\bold{1},\bold{1},\bold{1},\bold{5},\bold{1}] \right. \\ \nonumber & + & \left. \chi[\bold{1},\bold{1},\bold{1},\bold{1},\bold{5}] + 3\chi[\bold{3},\bold{1},\bold{1},\bold{1},\bold{1}]+2\chi[\bold{1},\bold{3},\bold{1},\bold{1},\bold{1}]+2\chi[\bold{1},\bold{1},\bold{3},\bold{1},\bold{1}]+2\chi[\bold{1},\bold{1},\bold{1},\bold{3},\bold{1}] \right. \\ \nonumber & + & \left. 2\chi[\bold{1},\bold{1},\bold{1},\bold{1},\bold{3}] + 8 + \chi[\bold{3},\bold{3},\bold{1},\bold{1},\bold{1}]+\chi[\bold{3},\bold{1},\bold{3},\bold{1},\bold{1}]+\chi[\bold{3},\bold{1},\bold{1},\bold{3},\bold{1}]+\chi[\bold{3},\bold{1},\bold{1},\bold{1},\bold{3}] \right. \\ \nonumber & + & \left. \chi[\bold{1},\bold{3},\bold{3},\bold{1},\bold{1}]+\chi[\bold{1},\bold{3},\bold{1},\bold{3},\bold{1}]+\chi[\bold{1},\bold{3},\bold{1},\bold{1},\bold{3}]+\chi[\bold{1},\bold{1},\bold{3},\bold{3},\bold{1}] \right. \\ \nonumber & + & \left. \chi[\bold{1},\bold{1},\bold{3},\bold{1},\bold{3}] + \chi[\bold{1},\bold{1},\bold{1},\bold{3},\bold{3}] + \chi[\bold{1},\bold{2},\bold{2},\bold{2},\bold{2}] \right. \\ \nonumber & + & \left. (a\sqrt{q t} + \frac{1}{a\sqrt{q t}})\left( \chi[\bold{2},\bold{1},\bold{2},\bold{2},\bold{1}] + \chi[\bold{2},\bold{2},\bold{1},\bold{1},\bold{2}] \right) \right. \\ \nonumber & + & \left. (b^2 + \frac{1}{b^2})\left( \chi[\bold{1},\bold{2},\bold{2},\bold{1},\bold{1}] + \chi[\bold{1},\bold{1},\bold{1},\bold{2},\bold{2}] \right) \right) + x^5 \left( \chi_y[\bold{4}](3+\chi[\bold{3},\bold{1},\bold{1},\bold{1},\bold{1}] \right. \\ \nonumber & + & \left. \chi[\bold{1},\bold{3},\bold{1},\bold{1},\bold{1}]+\chi[\bold{1},\bold{1},\bold{3},\bold{1},\bold{1}]+\chi[\bold{1},\bold{1},\bold{1},\bold{3},\bold{1}] + \chi[\bold{1},\bold{1},\bold{1},\bold{1},\bold{3}]) + \chi_y[\bold{2}] \left(\chi[\bold{5},\bold{1},\bold{1},\bold{1},\bold{1}] \right. \right. \\ \nonumber & + & \left. \left. \chi[\bold{1},\bold{5},\bold{1},\bold{1},\bold{1}]+\chi[\bold{1},\bold{1},\bold{5},\bold{1},\bold{1}]+\chi[\bold{1},\bold{1},\bold{1},\bold{5},\bold{1}] + \chi[\bold{1},\bold{1},\bold{1},\bold{1},\bold{5}] + 7\chi[\bold{3},\bold{1},\bold{1},\bold{1},\bold{1}] \right. \right. \\ \nonumber & + & \left. \left. 6\chi[\bold{1},\bold{3},\bold{1},\bold{1},\bold{1}]+6\chi[\bold{1},\bold{1},\bold{3},\bold{1},\bold{1}]+6\chi[\bold{1},\bold{1},\bold{1},\bold{3},\bold{1}] + 6\chi[\bold{1},\bold{1},\bold{1},\bold{1},\bold{3}] + 12 \right. \right. \\ \nonumber & + & \left. \left. 2\chi[\bold{3},\bold{3},\bold{1},\bold{1},\bold{1}]+2\chi[\bold{3},\bold{1},\bold{3},\bold{1},\bold{1}]+2\chi[\bold{3},\bold{1},\bold{1},\bold{3},\bold{1}]+2\chi[\bold{3},\bold{1},\bold{1},\bold{1},\bold{3}] + 2\chi[\bold{1},\bold{3},\bold{3},\bold{1},\bold{1}] \right. \right. \\ \nonumber & + & \left. \left. 2\chi[\bold{1},\bold{3},\bold{1},\bold{3},\bold{1}]+2\chi[\bold{1},\bold{3},\bold{1},\bold{1},\bold{3}]+2\chi[\bold{1},\bold{1},\bold{3},\bold{3},\bold{1}] + 2\chi[\bold{1},\bold{1},\bold{3},\bold{1},\bold{3}] + 2\chi[\bold{1},\bold{1},\bold{1},\bold{3},\bold{3}] \right. \right. \\ \nonumber & + & \left. \left. \chi[\bold{1},\bold{2},\bold{2},\bold{2},\bold{2}] + (a\sqrt{q t} + \frac{1}{a\sqrt{q t}})\left( \chi[\bold{2},\bold{1},\bold{2},\bold{2},\bold{1}] + \chi[\bold{2},\bold{2},\bold{1},\bold{1},\bold{2}] \right) \right. \right. \\ \nonumber & + & \left. \left. (b^2 + \frac{1}{b^2})\left( \chi[\bold{1},\bold{2},\bold{2},\bold{1},\bold{1}] + \chi[\bold{1},\bold{1},\bold{1},\bold{2},\bold{2}] \right) \right) \right)  + O(x^6)
\eea    
where we use $\chi_y[d]$ for the $d$ dimensional representation of $SU_y(2)$ and $\chi[d_1,d_2,d_3,d_4,d_5]$ for the representations of the appropriate dimensions under $SU(2)^5$ where the $SU(2)'s$ are ordered as $\chi[\sqrt{A},f,g,p,h]$.

\section{Spinor matter}

In this section we discuss the addition of matter in a spinor representation of an $SO$ gauge group. It is well known that there is no perturbative way to add matter in spinor representations through D-brane constructions in string theory. However, we claim that there is a way to do so for webs in the presence of $O5$-planes, in a non perturbative manner. We first present our conjecture for how this is done, and present our argument for why this gives spinor matter. We then proceed to give evidence for this conjecture.

\begin{figure}
\center
\includegraphics[width=0.6\textwidth]{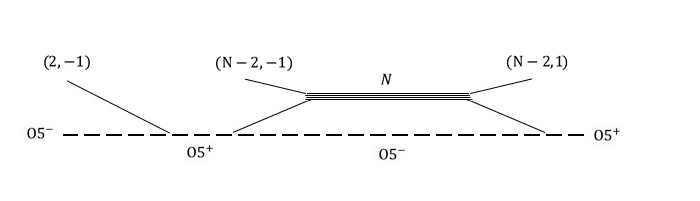} 
\caption{Adding a stuck NS5-brane to an $SO(2N)$ web.}
\label{XX1}
\end{figure}

\begin{figure}
\center
\includegraphics[width=0.6\textwidth]{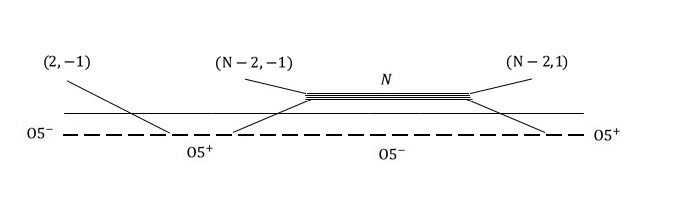} 
\caption{A web configuration describing an $SU(2)\times SO(2N+2)$ quiver with a half-bifundamental and one flavor for each group.}
\label{XX2}
\end{figure}

 We claim that the configuration shown in figure \ref{XX1}, in which we add a stuck NS5-brane to an $SO(2N)$ gauge theory, corresponds to adding a single hyper in the spinor representation of that group. Our motivation for this is as follows. First, consider the system in figure \ref{XX2}, describing an $1F+SU(2)\times SO(2N+2)+1F$ gauge theory. Starting from this system, we can get to the the one in figure \ref{XX1} by going on the Higgs branch described in the web by separating a full D5-brane. In the gauge theory this describes giving a vev to the operator $q B Q$ where $q$ is the $SU(2)$ fundamental, $B$ the half bifundamental, and $Q$ the vector of $SO(2N+2)$. 

Perturbativly, this completely breaks the $SU(2)$ gauge group. However, the web suggests that in this limit we do remain with additional degrees of freedom. Thus, these can only come from instantons of the $SU(2)$. It is well known that the $1$ instanton of $SU(2)$ with $N_f$ flavors carries charges in the spinor representation of $SO(2N_f)$. Therefore, we conjecture that the remaining state can be described by an hypermultiplet in the spinor of $SO(2N)$ whose origin is non-perturbative in the brane web. Although we have used $SO(2N)$ in this example, the same reasoning can also be carried out for the $SO(2N+1)$ case.

\begin{figure}
\center
\includegraphics[width=0.6\textwidth]{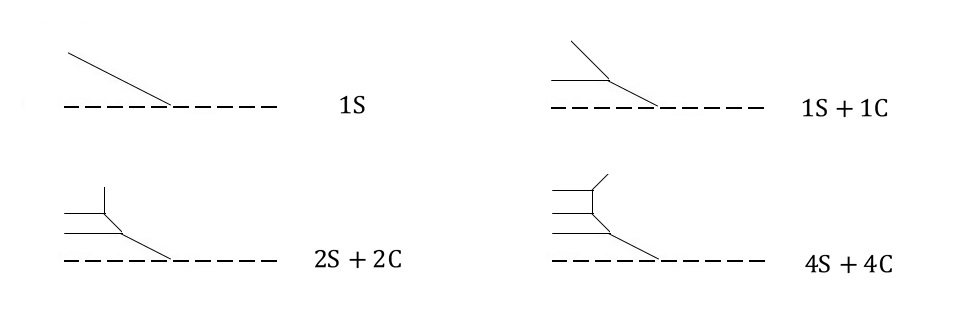} 
\caption{The matter contribution associated with a single stuck NS$5$-brane found using the preceding argument.}
\label{XX3}
\end{figure}

Before giving support for our claim, we wish to state some further implications of it. First, we can consider what happens when we attach further D$5$-branes as shown in figure \ref{XX3}. We can answer this question by again starting with the system of figure \ref{XX2} where adding the D$5$-brane corresponds to adding a flavor for the $SU(2)$. The instanton is again in the spinor of $SO(2N_f)$ which decomposes to two spinors of opposite chirality under $SO(2N_f-2)$. We thus conclude that we now get two hypermultiplets both spinors of $SO(2N)$, but of opposite chirality. Finally, we wish to consider what happens if we similarly add a $(2,1)$ $5$-brane in the other direction. Particularly, we still expect a spinor hyper, but we inquire whether it has the same chirality or not as it opposing friend. We can answer this by considering the appropriate equivalence of the system in \ref{XX2}, where the spinor should appear as the instanton of the $SU(2)$ gauge group. The chirality of this spinor is determined by the $SU(2)$ $\theta$ angle, and as this is identical in the two constructions, we conclude that the two spinors have the same chirality. To change the chirality between the two spinors, one would have to switch the $\theta$ angles of one of the $SU(2)$'s. As previously mentioned, we do not know if this can be done. 

\begin{figure}
\center
\includegraphics[width=1\textwidth]{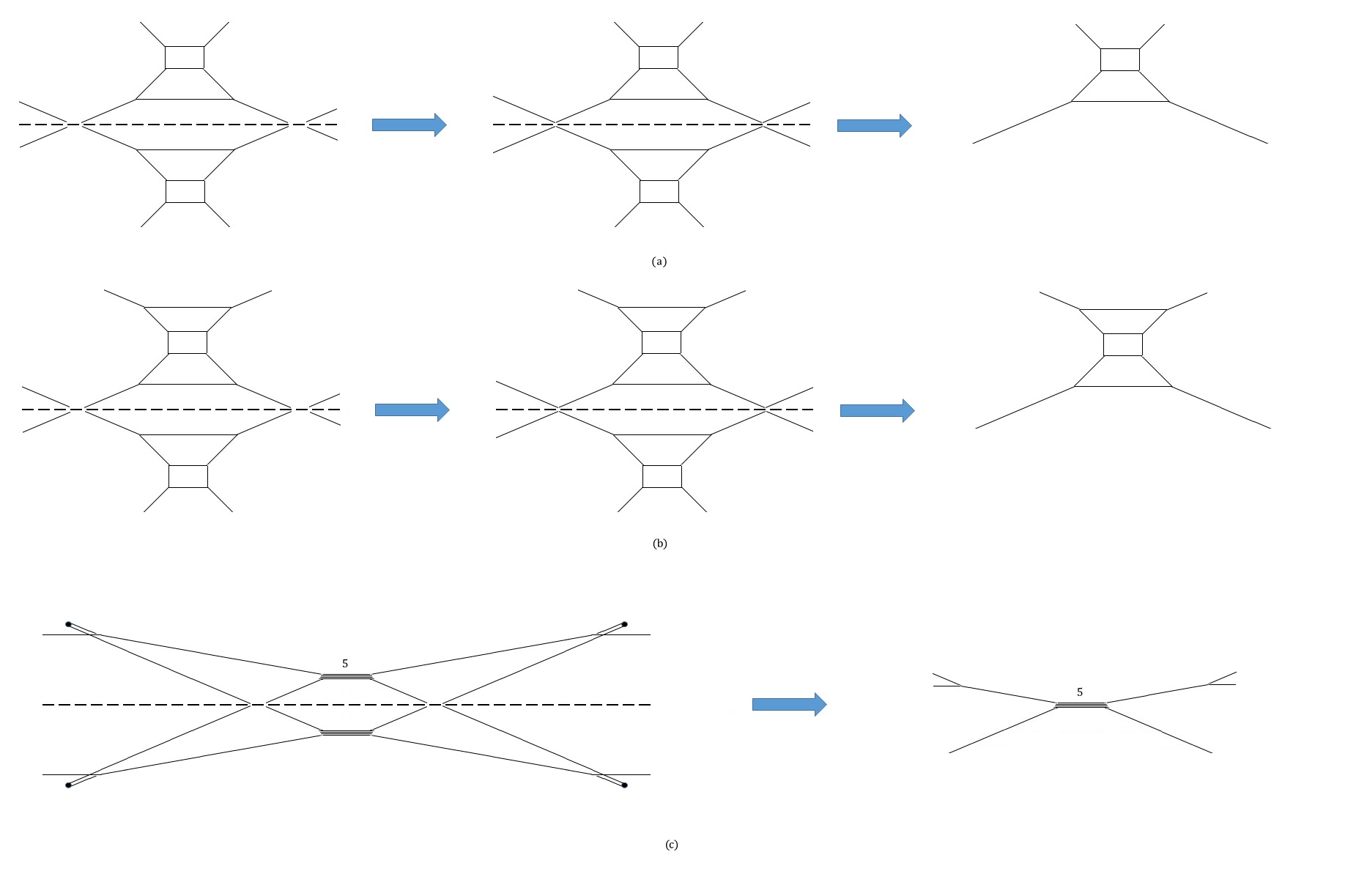} 
\caption{Several examples of the Higgs branch for theories of the form $SO(2N)+2S$ for $N=3,4,5$. Starting from the initial web on the right, we take the massless spinor limit, corresponding to taking the distance between the two pairs of NS5-branes to zero. Then a Higgs branch opens up given in the web by detaching the web from the orientifold. This is shown on the right where for ease of presentation we have shown only half the web. (a) The case of $N=3$. We know from the gauge theory that there is a Higgs branch breaking the theory to $SU(3)$. (b) The case of $N=4$. We know from the gauge theory that there is a Higgs branch breaking the theory to $SU(4)$. (c) The case of $N=5$. We know from the gauge theory that there is a Higgs branch breaking the theory to $SU(5)+2F$. In all $3$ cases the Higgs branch is correctly reproduced in the web.}
\label{XX4}
\end{figure}

\begin{figure}
\center
\includegraphics[width=1\textwidth]{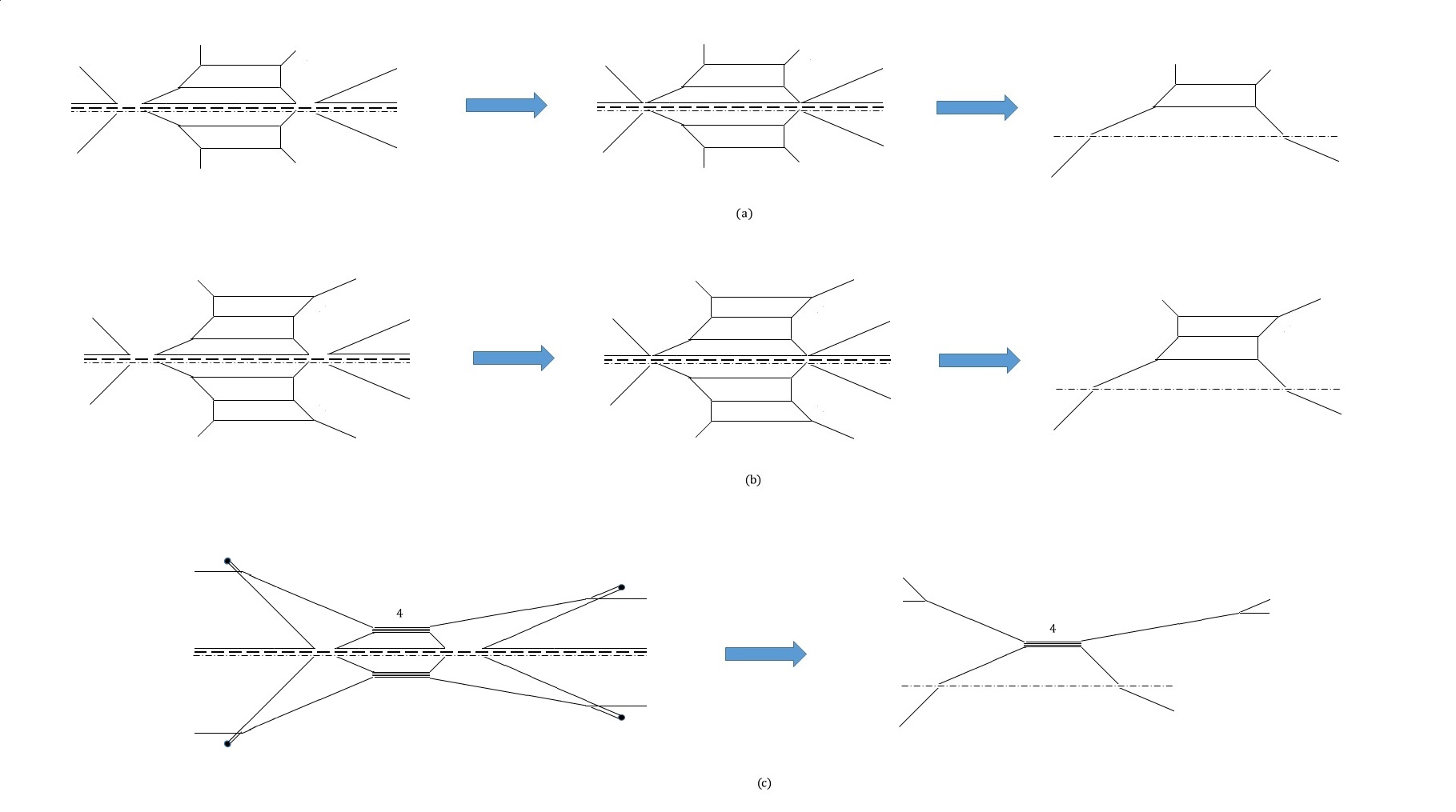} 
\caption{Several examples of the Higgs branch for theories of the form $SO(2N+1)+2S$ for $N=2,3,4$. Starting from the initial web on the right, we take the massless spinor limit, corresponding to taking the distance between the two pairs of NS5-branes to zero. Then a Higgs branch opens up given in the web by detaching the web from the orientifold. This is shown on the right where for ease of presentation we have shown only half the web. (a) The case of $N=2$. We know from the gauge theory that there is a Higgs branch breaking the theory to $SU(2)$. (b) The case of $N=3$. We know from the gauge theory that there is a Higgs branch breaking the theory to $SU(3)$. (c) The case of $N=4$. We know from the gauge theory that there is a Higgs branch breaking the theory to $SU(4)+2F$. In all $3$ cases the Higgs branch is correctly reproduced in the web.}
\label{XX5}
\end{figure}

We can now proceed to give evidence for our conjecture. First, we look at the Higgs branch. In figures \ref{XX4} and \ref{XX5} we show the webs for a variety of $SO$ groups with two spinors of the same chirality. These theories then have a Higgs branch breaking them to an $SU$ group. We show that the web correctly reproduces this branch, giving the expected theory. Furthermore, this branch can only be accessed when the spinor is effectively massless which in the web corresponds to the point where the would be $SU(2)$ instanton is massless as expected from our interpretation. 


\begin{figure}
\center
\includegraphics[width=1\textwidth]{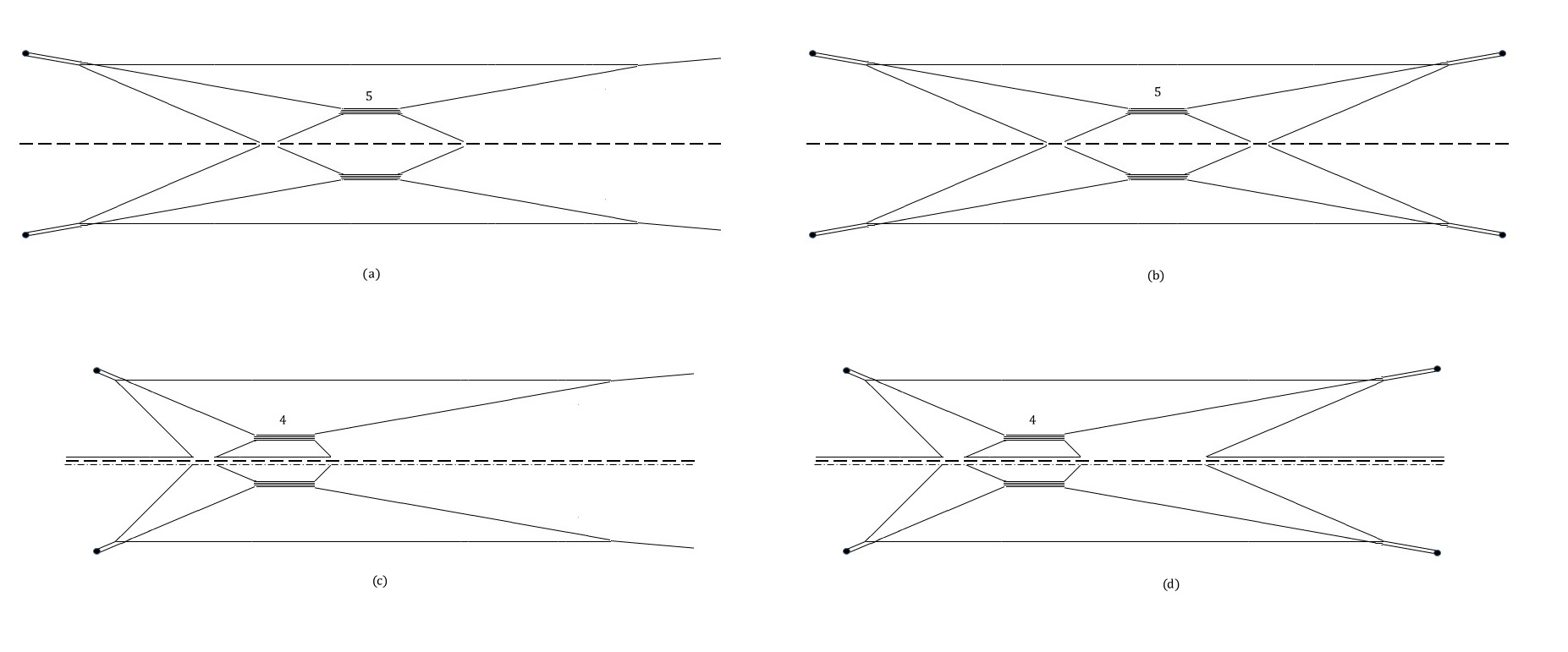} 
\caption{The webs we conjecture for (a) $SO(12)+\frac{1}{2}S$, (b) $SO(12)+1S$, (c) $SO(11)+\frac{1}{2}S$, (d) $SO(11)+1S$. One can see that the Higgs branch correctly agrees with the gauge theory expectation. The cases with a half-hyper, (a)+(c), do not have a Higgs branch. For (b) the gauge theory has a Higgs branch leading to $SU(6)$ which is correctly reproduced in the web. For (d) the gauge theory has a Higgs branch leading to $SU(5)$ which is again correctly reproduced in the web.}
\label{XX6}
\end{figure}

Note that we are essentially limited by the requirement that the would be $SU(2)$ gauge group sees less than $8$ flavors. Naively, this would imply the we cannot get spinors of $SO(N)$ for $N>10$. However, we can by a slight generalization get one also for $N=11,12$. In these theories, the spinors are pseudo-real, so a half-hyper is possible. Figure \ref{XX6} shows the webs we conjecture for $SO(11)$ and $SO(12)$ with one half or full hyper in the spinor representation. One can see that the Higgs branch of these webs agrees with what expected from the gauge theory. One issue with the webs for the full spinor cases is that they have only one mass deformation, in contrary to the two expected from the gauge theory. This is reminiscent of the web for $USp(2N)+AS$ which also has just one mass deformation. In that case the web is for a massless antisymmetric. Likewise this web appears to have a massless spinor.          

Using these webs we can look at the existence of fixed points for $SO$ groups with both vector and spinor matter. First, holding the spinor matter fixed, we can use the web to determine what is the maximal number of vectors one can add while still having a $5d$ fixed point. We generally find agreement with the expectations from \cite{Zaf1}. 

We can also ask what this implies about the limit of spinor matter. As mentioned, we are limited by the requirement that the would be $SU(2)$ gauge group sees less than $8$ flavors. Yet, we argue that this does not represent a limitation on spinor matter for $SO$ gauge theories, rather a breakdown of the interpretation of these webs. A limitation on the matter content due to a lack of $5d$ fixed point manifests as a lack of a brane web when taking the Yang-Mills coupling to infinity, generally due to intersecting legs that cannot be resolved by a finite number of HW transitions. This is not the case here, rather the intersection arises when we take the massless spinor limit indicating that there are in fact additional states in this case. Therefore, we do not think this gives a limit on spinor matter for $SO$ gauge theories, rather being a limitation of the method. 

\subsection{Dualities}

As our final piece of evidence, we examine dualities between systems involving $SO$ groups with spinor matter. The idea is to use the webs to motivate dualities and then test them using the superconformal index. This then provides independent evidence for the duality and thus also for the original identification leading to it. There is one limitation in this test as instanton counting for $SO$ groups with spinor matter is currently unknown. Thus, we are limited to comparing the perturbative parts.   

\subsubsection{Example $1$}

\begin{figure}
\center
\includegraphics[width=1.05\textwidth]{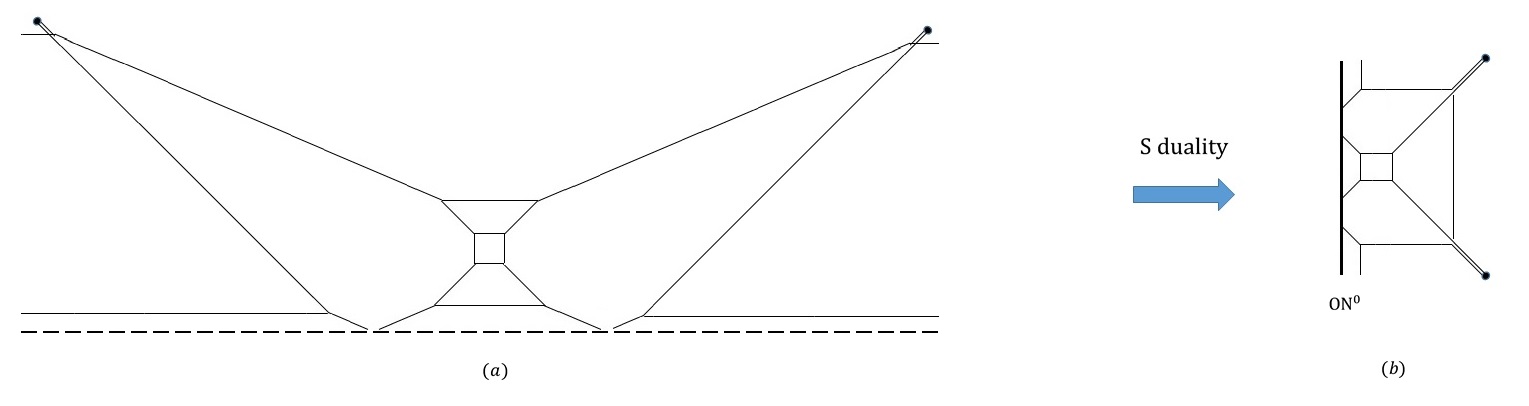} 
\caption{(a) The web for $SO(8)+2S+2C$. (b) The S-dual of the web in (a). It describes an $SU_0(2)\times USp_{\pi}(4)\times SU_0(2)$ gauge theory. The $USp_{\pi}(4)$ group can be seen by pulling the $(1,1)$ and $(1,-1)$ $7$-branes through the $5$-branes, and merging them to an $O7^-$ plane (see \cite{BZ1}).}
\label{XX7}
\end{figure}

As our first example, consider the theory shown in figure \ref{XX7} (a). From figure \ref{XX3}, we claim that this describes an $SO(8)+2S+2C$ gauge theory where we use $S$ and $C$ for the two Weyl spinor representations of $SO(8)$. We can take the S-dual description leading, as shown in figure \ref{XX7} (b), to the quiver theory $SU(2)\times USp(4)\times SU(2)$. Thus we conjecture that these two theories are dual. We want to also check this using the superconformal index. 

We start with the $SO(8)$ theory. The classical global symmetry is $U(1)\times USp(4)^2$ coming from the topological and flavor symmetries. There is also a $Z_2$ discrete symmetry coming from exchanging the two spinor representations. The analysis of \cite{Zaf1} suggests that there is no enhanced symmetry so it seems to also be the quantum symmetry. 

\begin{figure}
\center
\includegraphics[width=0.8\textwidth]{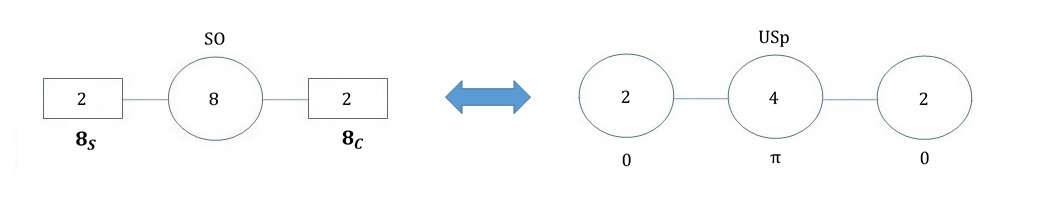} 
\caption{The duality suggested by the webs of figure \ref{XX7}.}
\label{fgr33}
\end{figure}

We preform the calculation up to order $x^5$ finding:

\bea
Index_{SO(8)} & = & 1 + x^2(1+\chi[\bold{10},\bold{1}] + \chi[\bold{1},\bold{10}]) + x^3 \chi_y[\bold{2}](2+\chi[\bold{10},\bold{1}] + \chi[\bold{1},\bold{10}]) \\ \nonumber & + & x^4 \left( \chi_y[\bold{3}](2+\chi[\bold{10},\bold{1}] + \chi[\bold{1},\bold{10}]) + \chi[\bold{35}_{(4,0)},\bold{1}] + \chi[\bold{1},\bold{35}_{(4,0)}] + \chi[\bold{14},\bold{1}]  \right. \\ \nonumber & + & \left. \chi[\bold{1},\bold{14}] + \chi[\bold{10},\bold{10}] + \chi[\bold{5},\bold{5}] + \chi[\bold{10},\bold{1}] + \chi[\bold{1},\bold{10}] + \chi[\bold{5},\bold{1}] + \chi[\bold{1},\bold{5}] + 3 \right)  \\ \nonumber & + &  x^5 \left( \chi_y[\bold{4}](2+\chi[\bold{10},\bold{1}] + \chi[\bold{1},\bold{10}]) + \chi_y[\bold{2}] (\chi[\bold{35}_{(4,0)},\bold{1}] + \chi[\bold{1},\bold{35}_{(4,0)}] + \chi[\bold{35}_{(2,1)},\bold{1}]  \right. \\ \nonumber & + & \left. \chi[\bold{1},\bold{35}_{(2,1)}] + \chi[\bold{14},\bold{1}] + \chi[\bold{1},\bold{14}] + 2\chi[\bold{10},\bold{10}] + \chi[\bold{5},\bold{5}] + 4\chi[\bold{10},\bold{1}] + 4\chi[\bold{1},\bold{10}]  \right. \\ \nonumber & + & \left. \chi[\bold{5},\bold{1}] + \chi[\bold{1},\bold{5}] + 4) \right) + O(x^6)
\eea
where we use $\chi[d_1,d_2]$ for the representation of dimension $d_1$ ($d_2$) under the first (second) $USp(4)$ symmetry. Since $USp(4)$ has two $35$ dimensional representations, both appearing in the index, we have also written their Cartan weight to distinguish between them. Note that the index is symmetric under the exchange of the two global $USp(4)$'s which is the manifestation of the discrete $Z_2$ symmetry.

Next we move to the $SU_0(2) \times USp_{\pi}(4) \times SU_0(2)$ theory. The classical global symmetry is $U(1)^3 \times SU(2)^2$, but as we will show this is enhanced at least to $U(1)\times USp(4)^2$ by the $SU(2)$'s $1$-instanton. There is also a discrete $Z_2$ symmetry of exchanging the two $SU(2)$ groups that matches the corresponding one in the $SO(8)$ theory. Next, we evaluate the index of this theory to order $x^5$. To that order we have contributions from the (1,0,0) + (0,0,1) + (1,0,1) + (2,0,0) + (0,0,2) - instantons. Using the fugacity spanning shown in figure \ref{fgr34}, we find:

\begin{figure}
\center
\includegraphics[width=0.4\textwidth]{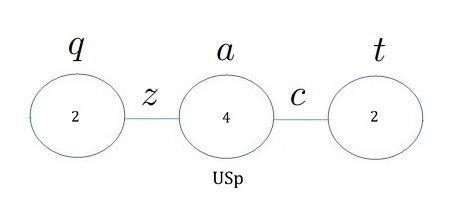} 
\caption{The fugacity spanning for the $SU_0(2) \times USp_{\pi}(4) \times SU_0(2)$ theory.}
\label{fgr34}
\end{figure}

\bea
Index_{SU^2(2)\times USp(4)} & = & 1 + x^2(5+ z^2 + \frac{1}{z^2} + c^2 + \frac{1}{c^2} + (q+\frac{1}{q})(z^2 + 1 + \frac{1}{z^2}) \nonumber \\  & + & (t+\frac{1}{t})(c^2 + 1 + \frac{1}{c^2})) + x^3 \chi_y[\bold{2}](6+ z^2 + \frac{1}{z^2} + c^2 + \frac{1}{c^2} \nonumber \\  & + & (q+\frac{1}{q})(z^2 + 1 + \frac{1}{z^2}) + (t+\frac{1}{t})(c^2 + 1 + \frac{1}{c^2}))  \nonumber \\  & + & O(x^4)
\eea

Although we evaluated the index to order $x^5$, we have written it only up to $x^3$ to avoid over-clutter. From the $x^2$ terms one can see the conserved currents of the classical global symmetry as well as ones provided by the (1,0,0) + (0,0,1) - instantons results in the enhancement of $U(1)^2 \times SU(2)^2 \rightarrow USp(4)^2$. This matches the global symmetries of the two theories and one can see that also the indices match to order $x^3$. We have also confirmed that the matching persists up to order $x^5$.

\subsubsection{Example $2$}

\begin{figure}
\center
\includegraphics[width=0.8\textwidth]{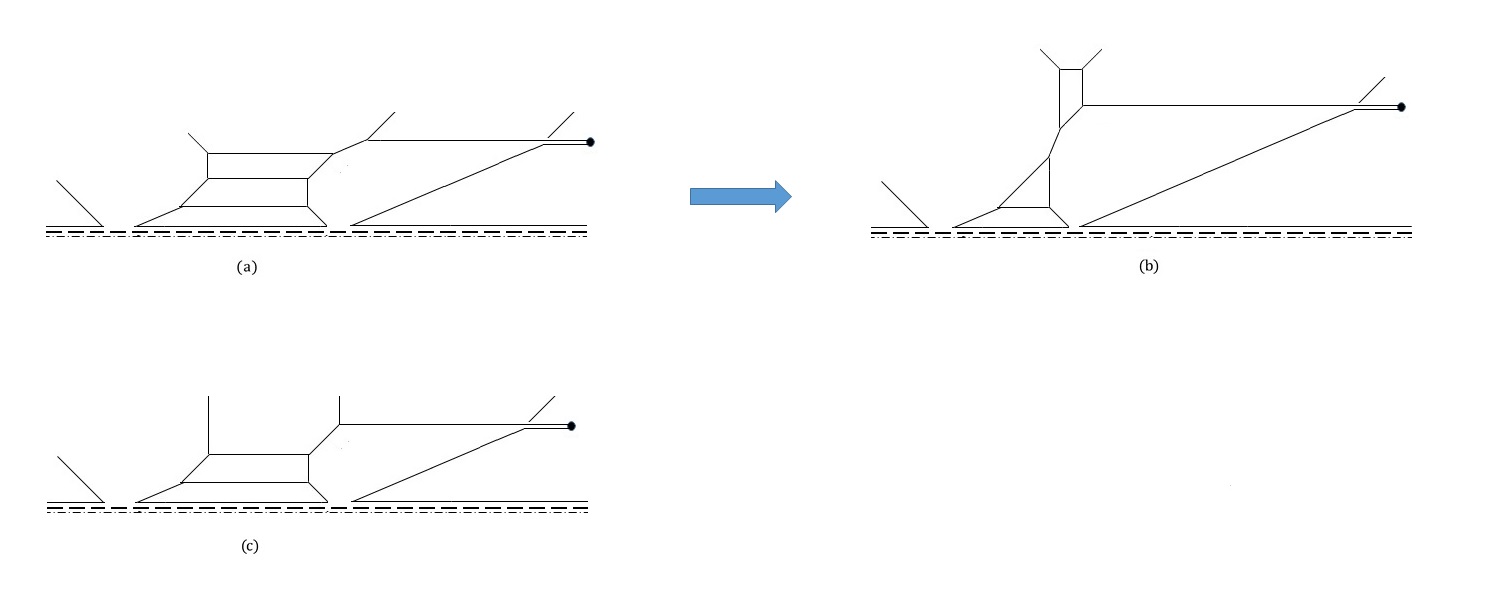} 
\caption{(a) The web for $SO(7)+1V+2S$. (b) Mass deforming the web of (a) leads to this web. One can see that it describes an $SU(2)$ gauging of the web in (c). (c) The web for $USp(4)+1AS+2F$, where we have used that $SO(5)=USp(4)$ under which the fundamental and antisymmetric representations of $USp(4)$ are the spinor and vector representations of $SO(5)$ respectively.}
\label{XX8}
\end{figure}

Our next examples involves $SO(N)$ groups with $N$ odd. As the webs for these theories involves an $\tilde{O5}^-$ plane with a stuck monodromy, performing S-duality is difficult. However, we can still overcome this by simply considering the guage theory on the NS$5$-branes. For example, consider the web of figure \ref{XX8} (a) describing $SO(7)+1V+2S$. We can mass deform it as shown in figure \ref{XX8} (b). From the point of view of the NS$5$-branes, this describes an $SU_0(2)$ gauging of the SCFT described by $USp(4)+1AS+2F$ (see figure \ref{XX8} (c)). The gauging is done into the topological symmetry of the $USp(4)$ gauge theory, but as this theory has an enhancement of symmetry to $SU(3)$ \cite{SMI}, we can rotate it so as to sit in the flavor sector.     

So we conclude that the dual is  $AS+USp(4) \times SU_0(2)$ where all that's left is to determine the $\theta$ angle of $USp(4)$. Comparing global symmetries, we see that they match: the $SO(7)$ theory having a $U_T(1)\times SU_V(2)\times USp_S(4)$ global symmetry while the quiver having a $U_T(1)\times SU_{AS}(2)\times USp(4)$ one (here we have used the enhancement of $U(1)\times SU_{BF}(2)\rightarrow USp(4)$ coming from the $1$-instanton of the gauge $SU(2)$ seen in the previous example). We thus see that the dual must be $AS+USp_{\pi}(4) \times SU_0(2)$, as otherwise there would be an additional enhancement not expected in the $SO(7)$ theory\footnote{Again, the results of \cite{Zaf1} suggests no enhancement in this case.}.  

We want to further test this duality by comparing the superconformal index. There are two problems with this calculation. One, due to the presence of the spinor matter, we cannot calculate the $SO(7)$ instanton contribution so we can only calculate the perturbative part. Two, there is a problem calculating di-group instantons in the $AS+USp_{\pi}(4) \times SU_0(2)$ theory. Calculating instantons in $USp+AS$ requires removing decoupled states where the full instanton partition function, $\mathcal{Z}$, contains extraneous contributions that must be removed by hand. This case is well understood, and the form of the decoupled states was worked out in \cite{HKKP}. Defining $\mathcal{Z}_c$ for the full instanton partition function with the extraneous contributions removed, we find:

\be
\mathcal{Z}_c = \mathcal{Z} PE\left[\frac{a x^2 \chi_{SU(2)}[\bold{2}]}{(1-x y)(1-\frac{x}{y})(1-x c)(1-\frac{x}{c})}\right] \label{erererex}
\ee 
where $c$ stands for the antisymmetric $SU(2)$ fugacity, $a$ for the $USp(4)$ instanton fugacity, and $\chi_{SU(2)}[\bold{2}]$ is the character for the fundamental of the gauge $SU(2)$ (which is a global symmetry from the $USp(4)$ point of view). We also use $PE[x]$ for the plethystic exponent of $x$, and this term in (\ref{erererex}) gives the contribution of the decoupled states. One notes that they carry gauge charges under the $SU(2)$ gauge group. These are responsible for the lack of enhancement, but also imply non-trivial interaction between these decoupled states and the $SU(2)$ gauge group degrees of freedom. Therefore, while $\mathcal{Z}_c$ should properly capture $USp(4)$ or $SU(2)$ instantons, we expect additional extraneous contributions for di-group instantons making these calculations unreliable. 

Bearing this in mind, we next state our result. We start with the $SO(7)+1V+2S$ theory. The classical global symmetry consists of a topological $U(1)$, an $SU(2)$ associated with the vector and a $USp(4)$ associated with the $2$ spinors. We calculate the perturbative index to order $x^5$ finding:

\bea
Index_{SO(7)} & = & 1 + x^2(1+\chi[\bold{3},\bold{1}] + \chi[\bold{1},\bold{10}]) + x^3 \left( \chi_y[\bold{2}](2+\chi[\bold{3},\bold{1}] + \chi[\bold{1},\bold{10}])  \right. \nonumber \\  & + & \left. \chi[\bold{2},\bold{1}] + \chi[\bold{2},\bold{5}] \right) + x^4 \left( \chi_y[\bold{3}](2+\chi[\bold{3},\bold{1}] + \chi[\bold{1},\bold{10}]) + \chi_y[\bold{2}](\chi[\bold{2},\bold{1}] + \chi[\bold{2},\bold{5}]) \right. \nonumber \\  & + & \left. \chi[\bold{1},\bold{35}_{(4,0)}] + \chi[\bold{1},\bold{14}] + \chi[\bold{3},\bold{10}] + \chi[\bold{1},\bold{10}] + \chi[\bold{1},\bold{5}] + \chi[\bold{5},\bold{1}] + \chi[\bold{3},\bold{1}] + 4 \right) \nonumber \\  & + &  x^5 \left( \chi_y[\bold{4}](2+\chi[\bold{3},\bold{1}] + \chi[\bold{1},\bold{10}]) + \chi_y[\bold{3}](\chi[\bold{2},\bold{1}] + \chi[\bold{2},\bold{5}]) \right. \nonumber \\  & + & \left. \chi_y[\bold{2}] (\chi[\bold{1},\bold{35}_{(4,0)}] + \chi[\bold{1},\bold{35}_{(2,1)}] + \chi[\bold{1},\bold{14}] + 2\chi[\bold{3},\bold{10}] + 4\chi[\bold{1},\bold{10}] + \chi[\bold{1},\bold{5}] \right. \nonumber \\  & + & \left. \chi[\bold{5},\bold{1}] + 4\chi[\bold{3},\bold{1}] + 5) + \chi[\bold{4},\bold{1}] + \chi[\bold{4},\bold{5}] + \chi[\bold{2},\bold{35}_{(2,1)}] + \chi[\bold{2},\bold{10}] + 2\chi[\bold{2},\bold{5}] \right. \nonumber \\  & + & \left. \chi[\bold{2},\bold{1}] \right) + O(x^6)
\eea 

Now we wish to compare this to the index of $AS+USp_{\pi}(4) \times SU_0(2)$. We continue to use $z, q$ for the fugacities of the gauge $SU(2)$ bifundamental and topological symmetries while the rest of the fugacities are as in (\ref{erererex}). To order $x^5$ we get contributions from the (0,1) + (0,2) + (1,0) + (2,0) instantons, where only the (2,0) instantons contribute states charged under the $USp(4)$ topological $U(1)$ (the (1,0) instantons are gauge-charged and only contribute through an invariant with the anti-instanton). We first separate them out, since we expect these states to match the instantons of $SO(7)$. For the others we find:

\bea
Index_{AS+USp(4)\times SU(2)} & = & 1 + x^2(4+ z^2 + \frac{1}{z^2} + c^2 + \frac{1}{c^2} + (q+\frac{1}{q})(z^2 + 1 + \frac{1}{z^2})) \nonumber \\  & + & x^3 \left( \chi_y[\bold{2}](5+ z^2 + \frac{1}{z^2} + c^2 + \frac{1}{c^2} + (q+\frac{1}{q})(z^2 + 1 + \frac{1}{z^2})) \right. \nonumber \\  & + & \left. (c+\frac{1}{c})(2+ q + \frac{1}{q} + z^2 + \frac{1}{z^2})  \right) + O(x^4)
\eea  

One can see that the two indices match, and, indeed, we have checked that they match up to order $x^5$.

Finally, we can consider the contributions of states charged under the $USp(4)$ topological $U(1)$. To order $x^5$, the only contributions we find come from the (2,0) instanton where we get:

\bea
Index^{(0,2)}_{AS+USp(4)\times SU(2)} & = & x^5 (a^2+\frac{1}{a^2})(c+\frac{1}{c}) + O(x^6) \label{fgfg}
\eea

We expect this to match against instanton contribution of $SO(7)+1V+2S$, but unfortunately we cannot verify it by direct calculation. 

\subsubsection{Example $3$}

\begin{figure}
\center
\includegraphics[width=0.7\textwidth]{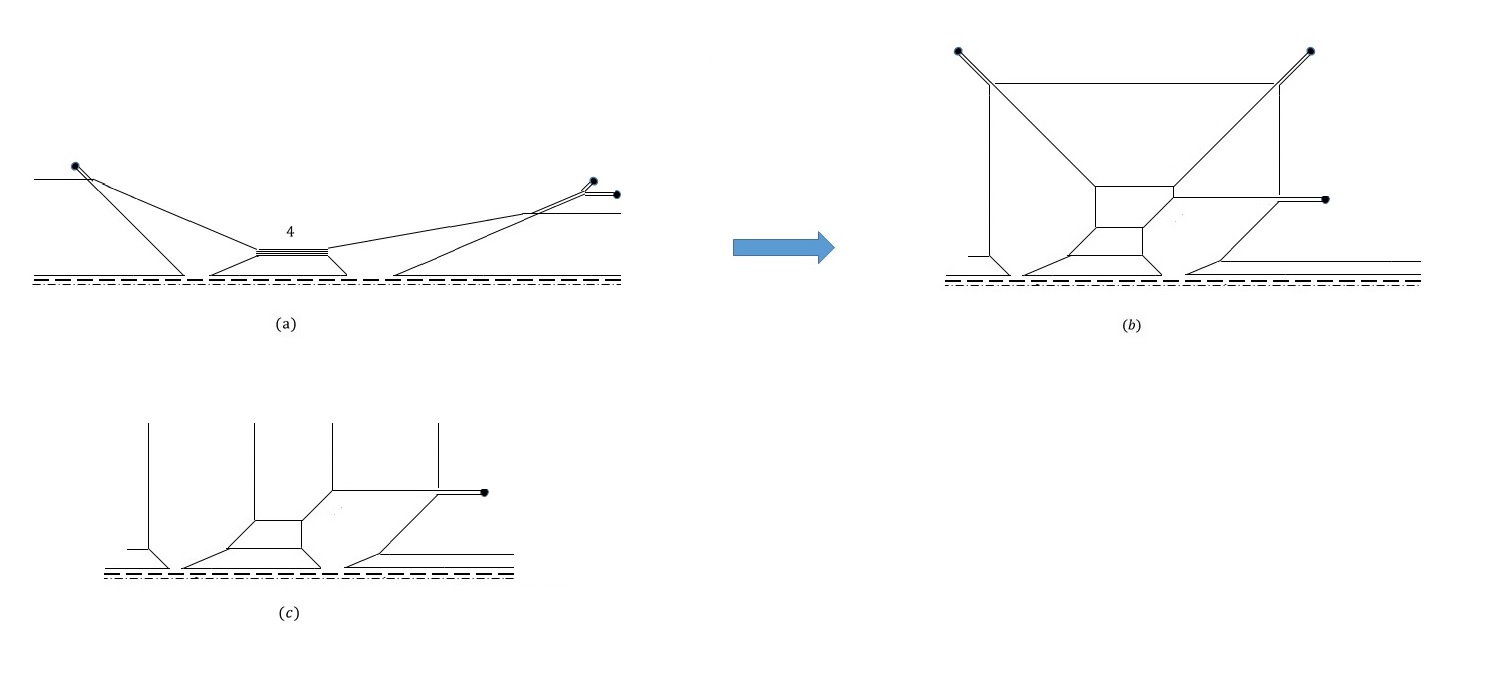} 
\caption{(a) The web for $SO(9)+1V+2S$. (b) Mass deforming the web of (a) leads to this web which describes a $USp(4)$ gauging of the web describing $USp(4)+1AS+4F$, shown in (c). That the gauging is done by a $USp(4)$ group can again be seen by pulling the $(1,1)$ and $(1,-1)$ $7$-branes through the two D$5$-branes. The resulting pair of $7$-branes can be interpreted as the S-dual of a resolved $O7^-$-plane.}
\label{XX9}
\end{figure}

As our final example, we consider the gauge theory $SO(9)+1V+2S$. By arguments similar to the previous ones, we conjecture that the dual should be $USp_{\pi}(4) \times USp_0(4)+AS$. The brane web for this theory is shown in figure \ref{XX9} (a). We can mass deform it to the web of \ref{XX9} (b). Looking from the NS$5$-branes point of view one can see that it describes a $USp_{\pi}(4)$ gauging of the SCFT described by $USp(4)+1AS+4F$. Since the gauged symmetry is realized on the NS$5$-branes, it is instantonic from the D$5$-branes point of view. However the $USp(4)+1AS+4F$ theory as an enhancement of $U_T(1)\times SO_F(8)\rightarrow SO(10)$\cite{SMI} which we can use to rotate the gauging to the flavor symmetry. Thus, we conclude that the dual is $USp_{\pi}(4) \times USp_0(4)+AS$ where the last $\theta$ angle was chosen so that the global symmetries match. 

Particularly, the $SO(9)$ theory as a classical symmetry given by $U_T(1)\times SU_V(2)\times USp_S(4)$ while the $USp$ quiver as a classical $U_T(1)^2 \times SU_{BF}(2)\times SU_{AS}(2)$ global symmetry. However we find that when $\theta=0$, for the $USp$ group with the antisymmetric, there is an additional enhancement of $U_T(1)\times SU_{BF}(2)\rightarrow USp(4)$. Thus, with this choice of $\theta$ angle, the global symmetries of the two theories agree again up to additional enhancements on either side\footnote{Like in the previous cases the analysis in \cite{Zaf1} suggests no enhancement for the $SO(9)$ theory.}.  

We next test this by calculating and comparing the superconformal index. There are two major limitations in this calculation. First we cannot calculate the $SO(9)$ instanton contribution due to the presence of spinor matter. Second, we cannot reliably calculate di-group instantons for the $USp^2$ theory. The reasons are the same as before: instanton counting for $USp(4)+AS$ requires removing the contributions of decoupled states. The precise form of these decoupled states was worked out in \cite{HKKP}, and in our case the removal is done by:

\be
\mathcal{Z}_c = \mathcal{Z} PE\left[\frac{q x^2 (z^2 + 1 + \frac{1}{z^2} + \chi_{USp(4)}[\bold{5}])}{(1-x y)(1-\frac{x}{y})(1-x c)(1-\frac{x}{c})}\right] \label{ererexx}
\ee 
where $c$ is the antisymmetric fugacity, $z$ the bifundamental, $q$ the topological one for the group with the antisymmetric and $\chi_{USp(4)}[\bold{5}]$ is the character for the antisymmetric of the gauge $USp_{\pi}(4)$ (which is a global symmetry from the $USp(4)+AS$ point of view). Again The gauge charges imply additional corrections for di-group instantons are expected so (\ref{ererexx}) is insufficient for the evaluation of di-group instanton contributions.

Finally, we state our results. For the perturbative index of $SO(9)+1V+2S$ we find:

\bea
Index_{SO(9)} & = & 1 + x^2(1+\chi[\bold{3},\bold{1}] + \chi[\bold{1},\bold{10}]) + x^3 \left( \chi_y[\bold{2}](2+\chi[\bold{3},\bold{1}] + \chi[\bold{1},\bold{10}])  \right.  \\ \nonumber & + & \left. \chi[\bold{2},\bold{10}] \right) + x^4 \left( \chi_y[\bold{3}](2+\chi[\bold{3},\bold{1}] + \chi[\bold{1},\bold{10}]) + \chi_y[\bold{2}]\chi[\bold{2},\bold{10}] + \chi[\bold{1},\bold{35}_{(4,0)}] \right.  \\ \nonumber & + & \left. 2\chi[\bold{1},\bold{14}] + \chi[\bold{3},\bold{10}] + \chi[\bold{1},\bold{10}] + 2\chi[\bold{1},\bold{5}] + \chi[\bold{5},\bold{1}] + \chi[\bold{3},\bold{1}] + 4 \right) + O(x^5)
\eea 
where we are working to order $x^4$.

To this order, in the $USp_{\pi}(4) \times USp_0(4)+AS$ theory, we get contributions of the (0,1)+(0,2) instantons finding:

\bea
Index_{USp(4)^2+AS} & = & 1 + x^2(4+ z^2 + \frac{1}{z^2} + c^2 + \frac{1}{c^2} + (q+\frac{1}{q})(z^2 + 1 + \frac{1}{z^2})) \nonumber \\  & + & x^3 \left( \chi_y[\bold{2}](5+ z^2 + \frac{1}{z^2} + c^2 + \frac{1}{c^2} + (q+\frac{1}{q})(z^2 + 1 + \frac{1}{z^2})) \right. \nonumber \\  & + & \left. (c+\frac{1}{c})(2 + z^2 + \frac{1}{z^2} + (q+\frac{1}{q})(z^2 + 1 + \frac{1}{z^2}))  \right) \nonumber \\  & + & O(x^4)
\eea 
where the fugacities are the ones used in (\ref{ererexx}). One can see that the indices match to the order shown, and we have further checked that the $x^4$ order also matches. 

\section{Conclusions}

In this article we studied brane webs in the presence of an $O5$-plane. This supports the existence of a wide class of new fixed points, and can be used to further study various aspects of these theories, such as dualities and enhancement of symmetry. The gauge theories that can be constructed in this way include alternating linear quivers of $SO$ and $USp$ groups as well as $D$ shaped quivers of $SU$ groups.

 We have also argued that one can engineer $SO(N)$ groups with spinor matter, where the spinor matter is thought to arise non-perturbatively. We would like to see if further evidence can be found for this. It will be interesting to further study the gauge theory leaving on the D$1$-brane associated with the $SO(N)$ instanton. These gauge theories are known to play an important role in instanton counting, and so may lead to a better understanding of instanton counting for $SO$ groups with spinor matter, which is currently unknown.

Finally, when sufficient flavors are introduced a $5d$ gauge theory may go to a $6d$ $\mathcal{N}$$=(1,0)$ SCFT, instead of a $5d$ SCFT. A well known example is $SU(2)+8F$ which has the $6d$ rank $1$ E-string theory as its UV fixed point\cite{SMG}. These sort of relations have been studied extensively recently for theories with ordinary brane web representations\cite{HKLTY,Zaf2,HKLY,OS}. This phenomenon appears to also occur for some of the theories considered in this article, as seen for example by the apparent presence of affine global symmetries\cite{Zaf1}. It will be interesting if this can be better understood, and if the $6d$ $\mathcal{N}$$=(1,0)$ SCFT's that these theories go to can be uncovered.     

\subsection*{Acknowledgments}

I would like to thank Oren Bergman, Soek Kim, and Hee-Cheol Kim for useful comments and discussions. G.Z. is supported in part by the Israel Science Foundation under grant no. 352/13, and by the German-Israeli Foundation for Scientific Research and Development under grant no. 1156-124.7/2011.

\appendix

\section{Index computation}

This appendix provides a brief review of the $5d$ superconformal index, and it's calculation using localization. The superconformal index is a sum of the BPS operators of a theory where if two or more operators can merge to form a non-BPS multiplet, they sum to zero. This is a useful quantity as it is invariant under continuous deformations since the spectrum of BPS operators can only change via this merging.

Specifically for the case of $5d$ $\mathcal{N}=1$ SCFT, the representations of the superconformal group are labeled by the highest weight of its $SO_L(5) \times SU_R(2)$ subgroup. We will call the two weights of $SO_L(5)$ as $j_1, j_2$ and those of $SU_R(2)$ as $R$. Then following \cite{KKL} the index is: 

\be
\mathcal{I}={\rm Tr}\,(-1)^F\,x^{2\,(j_1+R)}\,y^{2\,j_2}\,\mathfrak{q}^{\mathfrak{Q}}\, \label{eq:ind}
\ee
where $x,\,y$ are the fugacities associated with the superconformal group, while the fugacities collectively denoted by $\mathfrak{q}$ correspond to other commuting charges $\mathfrak{Q}$, generally flavor and topological symmetries. For a $5d$ gauge theory the index can be evaluated by localization where it is given by\cite{KKL}:

\be
\mathcal{I}= \int d\alpha \mathcal{Z}_{pert} |\mathcal{Z}_{Nek}|^2
\ee
where the integral is over the Cartan subalgebra of the gauge group. The terms $\mathcal{Z}_{pert}$ and $\mathcal{Z}_{Nek}$ are the contributions of the perturbative and instanton sectors respectively. The perturbative contribution, $\mathcal{Z}_{pert}$, can be easily evaluated using the results of \cite{KKL}. The instanton contribution, also known as the $5d$ Nekrasov partition function\cite{NS}, is harder to evaluate. In general $\mathcal{Z}_{pert}$ is expanded in a power series in the instanton fugacity, each term providing the contribution of the associated instantons.

These terms can in turn be evaluated by a contour integral where the integrand receiving contributions from the various matter and gauge content of the theory. The expressions for most of these contributions that we need have appeared elsewhere, notably \cite{KKL,HKKP,BGZ,BZ,BZ1}, and we won't repeat them here. The only exception being the $SO \times USp$ bifundamental and half-bifundamental whose expressions we provide below. In addition one also has to supplement this with a pole prescription detailing which poles are inside the contour. A good review of these is given in \cite{HKKP}.

Finally, the evaluation of the Nekrasov partition function is sometimes plagued with the contributions of extraneous degrees of freedom that must be removed by hand. These can materialize in the partition function as a breakdown of $x \rightarrow \frac{1}{x}$ invariance, which must be obeyed as it is part of the superconformal algebra. Another way these can appear in the partition function is as an infinite tower with representations of increasing dimension under a flavor symmetry. Examples and ways of dealing with the former can be found in \cite{BMPTY,HKT,BGZ}, and those for the latter in \cite{KKL,HKKP}.

\subsection{$SO \times USp$}

In this subsection we state the contributions of the matter content to the Nekrasov partition function in the $SO \times USp$ formalism. The gauge contributions for both $SO$ and $USp$ groups were already written elsewhere so we will not restate them. We concentrate on the contributions of bifundamentals and half-bifundamentals. In 4d these were considered in \cite{HKS}. The 5d results in the $O^+$ sector can be derived by lifting the 4d ones, but for the $O^-$ sector one has to derive these directly using the methods in \cite{Shad}. 

We start with the contribution of a full $SO(M) \times USp(2N)$ bifundamental hyper to the integrand for the $(k,K)$ instanton. The dual gauge group in this case is $USp(2k)\times O(K)$. We shall employ the notation $M=2n_1 + \chi_1$ and $K=2n_2 + \chi_2$ where $\chi = 0,1$. The $O(K)$ group has two disconnected parts, denoted as the $O^+$ and $O^-$, which must both be taken into account. We also separate the $O^-$ case to two distinct cases depending on whether $k$ is even or odd. Throughout this subsection we use the fugacities: $z$ for the bifundamental $U(1)$, $a$ for the $SO(M)$ gauge symmetry, $b$ for the $USp(2N)$ gauge symmetry, $u$ for the $USp(2k)$ dual gauge group, and $v$ for the $O(K)$ dual gauge group. The complete contribution is:

\bea
& & Z^{SO\times USp}_{BF+} =  \left[\prod^{n_1}_{i=1} (z+\frac{1}{z}-a_i-\frac{1}{a_i}) \prod_{m=1}^{k} \frac{(z+\frac{1}{z}-u_m y-\frac{1}{u_m y})(z+\frac{1}{z}-\frac{u_m}{y}-\frac{y}{u_m})}{(z+\frac{1}{z}-u_m x-\frac{1}{u_m x})(z+\frac{1}{z}-\frac{u_m}{x}-\frac{x}{u_m})} \right]^{\chi_2} \\ \nonumber & & [\prod^{n_2}_{j=1} (z+\frac{1}{z}-v_j-\frac{1}{v_j}) ]^{\chi_1} [\sqrt{z}-\frac{1}{\sqrt{z}}]^{\chi_1 \chi_2} 
\prod_{i,j=1}^{n_1,n_2} (z+\frac{1}{z}-a_i v_j-\frac{1}{a_i v_j})(z+\frac{1}{z}-\frac{a_i}{v_j}-\frac{v_j}{a_i}) \\ \nonumber & & \prod_{n,m=1}^{N,k} (z+\frac{1}{z}-b_n u_m-\frac{1}{b_n u_m})(z+\frac{1}{z}-\frac{b_n}{u_m}-\frac{u_m}{b_n}) \\ \nonumber & & \prod_{m,j=1}^{k,n_2} \frac{(z+\frac{1}{z}-u_m v_j y-\frac{1}{u_m v_j y})(z+\frac{1}{z}-\frac{u_m v_j}{y}-\frac{y}{u_m v_j})(z+\frac{1}{z}-\frac{u_m}{v_j y}-\frac{v_j y}{u_m})(z+\frac{1}{z}-\frac{v_j}{y u_m}-\frac{y u_m}{v_j})}{(z+\frac{1}{z}-u_m v_j x-\frac{1}{u_m v_j x})(z+\frac{1}{z}-\frac{u_m v_j}{x}-\frac{x}{u_m v_j})(z+\frac{1}{z}-\frac{u_m}{v_j x}-\frac{v_j x}{u_m})(z+\frac{1}{z}-\frac{v_j}{x u_m}-\frac{x u_m}{v_j})} \label{ooo1}
\eea
for the $O^+$ part.

\bea
& & Z^{SO\times USp}_{BF-O} =  \prod^{n_1}_{i=1} (z+\frac{1}{z}+a_i+\frac{1}{a_i}) \prod_{m=1}^{k} \frac{(z+\frac{1}{z}+u_m y+\frac{1}{u_m y})(z+\frac{1}{z}+\frac{u_m}{y}+\frac{y}{u_m})}{(z+\frac{1}{z}+u_m x+\frac{1}{u_m x})(z+\frac{1}{z}+\frac{u_m}{x}+\frac{x}{u_m})} \\ \nonumber & & [(\sqrt{z}+\frac{1}{\sqrt{z}}) \prod^{n_2}_{j=1} (z+\frac{1}{z}-v_j-\frac{1}{v_j}) ]^{\chi_1}  
\prod_{i,j=1}^{n_1,n_2} (z+\frac{1}{z}-a_i v_j-\frac{1}{a_i v_j})(z+\frac{1}{z}-\frac{a_i}{v_j}-\frac{v_j}{a_i}) \\ \nonumber & & \prod_{n,m=1}^{N,k} (z+\frac{1}{z}-b_n u_m-\frac{1}{b_n u_m})(z+\frac{1}{z}-\frac{b_n}{u_m}-\frac{u_m}{b_n}) \\ \nonumber & & \prod_{m,j=1}^{k,n_2} \frac{(z+\frac{1}{z}-u_m v_j y-\frac{1}{u_m v_j y})(z+\frac{1}{z}-\frac{u_m v_j}{y}-\frac{y}{u_m v_j})(z+\frac{1}{z}-\frac{u_m}{v_j y}-\frac{v_j y}{u_m})(z+\frac{1}{z}-\frac{v_j}{y u_m}-\frac{y u_m}{v_j})}{(z+\frac{1}{z}-u_m v_j x-\frac{1}{u_m v_j x})(z+\frac{1}{z}-\frac{u_m v_j}{x}-\frac{x}{u_m v_j})(z+\frac{1}{z}-\frac{u_m}{v_j x}-\frac{v_j x}{u_m})(z+\frac{1}{z}-\frac{v_j}{x u_m}-\frac{x u_m}{v_j})} \label{ooo2}
\eea
for the $O^-$ part and odd $K$.

\bea
& & Z^{SO\times USp}_{BF-E} =  \prod^{n_1}_{i=1} (z^2+\frac{1}{z^2}-a^2_i-\frac{1}{a^2_i}) \prod_{m=1}^{k} \frac{(z^2+\frac{1}{z^2}-u^2_m y^2-\frac{1}{u^2_m y^2})(z^2+\frac{1}{z^2}-\frac{u^2_m}{y^2}-\frac{y^2}{u^2_m})}{(z^2+\frac{1}{z^2}-u^2_m x^2-\frac{1}{u^2_m x^2})(z^2+\frac{1}{z^2}-\frac{u^2_m}{x^2}-\frac{x^2}{u^2_m})} \\ \nonumber & & [(z-\frac{1}{z}) \prod^{n_2-1}_{j=1} (z+\frac{1}{z}-v_j-\frac{1}{v_j}) ]^{\chi_1}  
\prod_{i,j=1}^{n_1,n_2-1} (z+\frac{1}{z}-a_i v_j-\frac{1}{a_i v_j})(z+\frac{1}{z}-\frac{a_i}{v_j}-\frac{v_j}{a_i}) \\ \nonumber & & \prod_{n,m=1}^{N,k} (z+\frac{1}{z}-b_n u_m-\frac{1}{b_n u_m})(z+\frac{1}{z}-\frac{b_n}{u_m}-\frac{u_m}{b_n}) \\ \nonumber & & \prod_{m,j=1}^{k,n_2-1} \frac{(z+\frac{1}{z}-u_m v_j y-\frac{1}{u_m v_j y})(z+\frac{1}{z}-\frac{u_m v_j}{y}-\frac{y}{u_m v_j})(z+\frac{1}{z}-\frac{u_m}{v_j y}-\frac{v_j y}{u_m})(z+\frac{1}{z}-\frac{v_j}{y u_m}-\frac{y u_m}{v_j})}{(z+\frac{1}{z}-u_m v_j x-\frac{1}{u_m v_j x})(z+\frac{1}{z}-\frac{u_m v_j}{x}-\frac{x}{u_m v_j})(z+\frac{1}{z}-\frac{u_m}{v_j x}-\frac{v_j x}{u_m})(z+\frac{1}{z}-\frac{v_j}{x u_m}-\frac{x u_m}{v_j})} \label{ooo3}
\eea
for the $O^-$ part and even $K$.

The contributions of this bifundamental also add additional poles to the integrand. The prescription for dealing with them follows directly from the work of \cite{HKKP}. Doing the following redefinitions in the above expressions: $p=\frac{1}{z x}$ and $d=\frac{z}{x}$, the correct prescription is to assume $x, p, d << 1$ taking all the poles within the circles and reset $p=\frac{1}{z x}, d=\frac{z}{x}$ only at the end of the calculation.  

The generalization to half-bifundamentals follows straightforwardly, similarly to the 4d case done in \cite{HKS}. To avoid the need to add an half-fundamental, we specialize to the case $\chi_1=0$. When taking the limit of a massless half-bifundamental, that is $z\rightarrow 1$, the expressions (\ref{ooo1}-\ref{ooo3}) become a total square, and the expression squared is identified with the contribution of a half-bifundamental. Explicitly these are given by: 

\bea
& & Z^{SO\times USp}_{HBF+} =  \left[\prod^{n_1}_{i=1} (\sqrt{a_i}-\frac{1}{\sqrt{a_i}}) \prod_{m=1}^{k} \frac{(u_m+\frac{1}{u_m}- y-\frac{1}{y})}{(u_m+\frac{1}{u_m}- x-\frac{1}{x})} \right]^{\chi_2} \\ \nonumber & & \prod_{i,j=1}^{n_1,n_2} (a_i+\frac{1}{a_i}- v_j-\frac{1}{v_j}) \prod_{n,m=1}^{N,k} (u_m+\frac{1}{u_m}-b_n -\frac{1}{b_n})\\ \nonumber & & \prod_{m,j=1}^{k,n_2} \frac{(u_m+\frac{1}{u_m}-v_j y-\frac{1}{v_j y})(u_m+\frac{1}{u_m}-\frac{v_j}{y}-\frac{y}{v_j})}{(u_m+\frac{1}{u_m}-v_j x-\frac{1}{v_j x})(u_m+\frac{1}{u_m}-\frac{v_j}{x}-\frac{x}{v_j})} \label{oou1}
\eea
for the $O^+$ part.

\bea
& & Z^{SO\times USp}_{HBF-O} =  \prod^{n_1}_{i=1} (\sqrt{a_i}+\frac{1}{\sqrt{a_i}}) \prod_{m=1}^{k} \frac{(u_m+\frac{1}{u_m}+ y+\frac{1}{y})}{(u_m+\frac{1}{u_m}+ x+\frac{1}{x})} \\ \nonumber & & \prod_{i,j=1}^{n_1,n_2} (a_i+\frac{1}{a_i}- v_j-\frac{1}{v_j}) \prod_{n,m=1}^{N,k} (u_m+\frac{1}{u_m}-b_n -\frac{1}{b_n})\\ \nonumber & & \prod_{m,j=1}^{k,n_2} \frac{(u_m+\frac{1}{u_m}-v_j y-\frac{1}{v_j y})(u_m+\frac{1}{u_m}-\frac{v_j}{y}-\frac{y}{v_j})}{(u_m+\frac{1}{u_m}-v_j x-\frac{1}{v_j x})(u_m+\frac{1}{u_m}-\frac{v_j}{x}-\frac{x}{v_j})} \label{oou2}
\eea
for the $O^-$ part and odd $K$.

\bea
& & Z^{SO\times USp}_{HBF-E} =  \prod^{n_1}_{i=1} (a_i-\frac{1}{a_i}) \prod_{m=1}^{k} \frac{(u^2_m+\frac{1}{u^2_m}- y^2-\frac{1}{y^2})}{(u^2_m+\frac{1}{u^2_m}- x^2-\frac{1}{x^2})} \\ \nonumber & & \prod_{i,j=1}^{n_1,n_2-1} (a_i+\frac{1}{a_i}- v_j-\frac{1}{v_j}) \prod_{n,m=1}^{N,k} (u_m+\frac{1}{u_m}-b_n -\frac{1}{b_n})\\ \nonumber & & \prod_{m,j=1}^{k,n_2-1} \frac{(u_m+\frac{1}{u_m}-v_j y-\frac{1}{v_j y})(u_m+\frac{1}{u_m}-\frac{v_j}{y}-\frac{y}{v_j})}{(u_m+\frac{1}{u_m}-v_j x-\frac{1}{v_j x})(u_m+\frac{1}{u_m}-\frac{v_j}{x}-\frac{x}{v_j})} \label{oou3}
\eea
for the $O^-$ part and even $K$.

In some cases the contributions add additional poles to the integrand, and the prescription then follows from the previous case where one defines $p=\frac{1}{x}$ in the denominators of (\ref{oou1}-\ref{oou2}). The prescription is then to assume $x, p<<1$, and set $p=\frac{1}{x}$ only at the end of the calculation.


\begin{thebibliography}{40}

\bibitem{Dou}
  M.~R.~Douglas, 
   JHEP {\bf 1102}, 011 (2011)
  [arXiv:1012.2880 [hep-th]].

\bibitem{LPS}
  N.~Lambert, C.~Papageorgakis, and M.~Schmidt-Sommerfeld, 
   JHEP {\bf 1101}, 083 (2011)
  [arXiv:1012.2882 [hep-th]].

\bibitem{SEI}
  N.~Seiberg, 
   Phys.\ Lett.\  B388:753-760 (1996)
  [arXiv:9608111 [hep-th]].
  
\bibitem{SM}
  N.~Seiberg, D.~R.~Morrison,
   Nucl.\ Phys.\  B483:229-247 (1997)
  [arXiv:9609070 [hep-th]].

\bibitem{SMI}
  N.~Seiberg, D.~R.~Morrison and K.~Intriligator,
   Nucl.\ Phys.\  B497:56-100 (1997)
  [arXiv:9702198 [hep-th]].

\bibitem{HA}
  O.~Aharony, A.~Hanany,
   Nucl.\ Phys.\  B504:239-271 (1997)
  [arXiv:9704170 [hep-th]].
	
\bibitem{AHK}
  O.~Aharony, A.~Hanany, and B.~Kol,
   JHEP {\bf 9801}, 002 (1998)
  [arXiv:9710116 [hep-th]].

\bibitem{BPTY}
  L.~Bao, E.~Pomoni, M.~Taki, and F.~Yagi, 
  JHEP {\bf 1204}, 105 (2012)
  [arXiv:1112.5228 [hep-th]].

\bibitem{BGZ} 
  O.~Bergman, D.~Rodriguez-Gomez and G.~Zafrir,
	JHEP {\bf 1403}, 112 (2014)
  arXiv:1311.4199 [hep-th].

\bibitem{Zaf} 
  G.~Zafrir,
	JHEP {\bf 1412}, 116 (2014)
  arXiv:1408.4040 [hep-th].

\bibitem{BZ} 
  O.~Bergman, G.~Zafrir,
  arXiv:1410.2806 [hep-th].

\bibitem{BZ1}
  O.~Bergman and G.~Zafrir,
	[arXiv:1507.03860 [hep-th]].

\bibitem{DHIZ}
  O.~DeWolfe, T.~Hauer, A.~Iqbal, and B.~Zwiebach,
	Adv.\ Theor.\ Math.\ Phys.\ 3:1785-1833 (1999)
  [arXiv:9812028 [hep-th]].

\bibitem{DHIZ1}
  O.~DeWolfe, T.~Hauer, A.~Iqbal, and B.~Zwiebach,
	Adv.\ Theor.\ Math.\ Phys.\ 3:1835-1891 (1999)
  [arXiv:9812209 [hep-th]].
	
\bibitem{IV} 
  A.~Iqbal and C.~Vafa,
	Phys.\ Rev.\ D90  10, 105031 (2014)
  arXiv:1210.3605 [hep-th].

\bibitem{KKL}
  H.~-C.~Kim, S.~Kim, and K.~Lee,
  JHEP {\bf 1210}, 142 (2012) 
  [arXiv:1206.6781 [hep-th]].

\bibitem{BMPTY}
  L.~Bao, V.~Mitev, E.~Pomoni, M.~Taki and F.~Yagi,
	JHEP {\bf 1401}, 175 (2014)
  [arXiv:1310.3841 [hep-th]].
  
\bibitem{HKT}
  H.~Hayashi, H.~-C.~Kim and T.~Nishinaka,
	JHEP {\bf 1406}, 014 (2014)
  [arXiv:1310.3854 [hep-th]].

\bibitem{BB}
  F.~Benini, S.~Benvenuti, and Y.~Tachikawa,
  JHEP {\bf 0909}, 052 (2009)
  [arXiv:0906.0359 [hep-th]].

\bibitem{KB}
  I.~Brunner, A.~Karch,
   Phys.\ Lett.\  B409:109-116 (1997)
  [arXiv:9705022 [hep-th]].

\bibitem{HK}
  A.~Hanany, B.~Kol,
   JHEP {\bf 0006}, 013 (2000)
  [arXiv:0003025 [hep-th]].

\bibitem{CHFM}
  S.~Cremonesi, G.~Ferlito, A.~Hanany and N.~Mekareeya, 
  [arXiv:1505.06302 [hep-th]].

\bibitem{Hori}
  K.~Hori, 
  Nucl.\ Phys.\  B539:35-78 (1999)
	[arXiv:9805141 [hep-th]].
 
\bibitem{Tachi}
  Y.~Tachikawa, 
  JHEP {\bf 1111}, 123 (2011)
	[arXiv:1110.0531 [hep-th]].

\bibitem{Wit}
  E.~Witten, 
  Phys.\ Lett.\ B117:324-328 (1982)

\bibitem{HZ}
  A.~Hanany, A.~Zaffaroni,
   JHEP {\bf 9907}, 009 (1999)
  [arXiv:9903242 [hep-th]].

\bibitem{Tachi1.5}
  Y.~Tachikawa,
  [arXiv:1501.01031 [hep-th]].
	
\bibitem{Yon} 
  K.~Yonekura,
  [arXiv:1505.04743 [hep-th]].

\bibitem{Zaf1} 
  G.~Zafrir,
	JHEP {\bf 1507}, 087 (2015)
  arXiv:1503.08136 [hep-th].

\bibitem{HKKP}
  C.~Hwang, J.~Kim, S.~Kim and J.~Park, 
  [arXiv:1406.6793 [hep-th]].








\bibitem{SMG}
  O.~J.~Ganor, D.~R.~Morrison, and N.~Seiberg,
  Nucl.\ Phys.\  B487:93-127 (1997) 
  [arXiv:9610251 [hep-th]].

\bibitem{Zaf2} 
  G.~Zafrir,
  arXiv:1509.02016 [hep-th].











	


\bibitem{HKLTY} 
  H.~Hayashi, S.~Kim, K.~Lee, M.~Taki, and F.~Yagi,
  [arXiv:1505.04439 [hep-th]].

\bibitem{HKLY} 
  H.~Hayashi, S.~Kim, K.~Lee, and F.~Yagi,
  [arXiv:1509.03300 [hep-th]].

\bibitem{OS} 
  K.~Ohmori, H.~Shimizu,
  [arXiv:1509.03195 [hep-th]].

\bibitem{NS}
  N.~Nekrasov, S.~Shadchin, 
  Commun.\ Math.\ Phys.\ 252:359-391 (2004)
  [arXiv:0404225 [hep-th]].

\bibitem{HKS}
  L.~Hollands, C.~A.~Keller, and J.~Song,
  JHEP {\bf 1103}, 053 (2011) 
  [arXiv:1012.4468 [hep-th]].

\bibitem{Shad}
  S.~Shadchin, 
  [arXiv:0502180 [hep-th]].




\end{thebibliography}
\end{document}